\documentclass[aps,prd,floatfix,onecolumn,notitlepage,superscriptaddress,preprintnumbers,nofootinbib,10pt]{revtex4-1}

\usepackage{natbib,setspace}
\usepackage{graphicx}
\usepackage{amssymb}
\usepackage{amsmath,amssymb}
\usepackage{color}
\usepackage{fancyhdr}
\definecolor{darkblue}{rgb}{0,0,0.5}
\usepackage[colorlinks,linkcolor=blue,citecolor=blue,urlcolor=blue]{hyperref}

\allowdisplaybreaks

\begin{document}
\date{\today}

\title{Collider Constraints on a Dark Matter Interpretation of the XENON1T Excess}

\author{R. Primulando}
\email{rprimulando@unpar.ac.id}
\affiliation{Center for Theoretical Physics, Department of Physics, Parahyangan Catholic University, Jl. Ciumbuleuit 94, Bandung 40141, Indonesia}

\author{J. Julio} 
\email{julio@lipi.go.id}
\affiliation{Indonesian Institute of Sciences (LIPI), Kompleks Puspiptek Serpong, Tangerang 15314, Indonesia}

\author{P. Uttayarat}
\email{patipan@g.swu.ac.th}
\affiliation{Department of Physics, Srinakharinwirot University, 114 Sukhumvit 23 Rd., Wattana, Bangkok
10110, Thailand}
\affiliation{National Astronomical Research Institute of Thailand, Chiang Mai 50180, Thailand}

\begin{abstract}
In light of the excess in the low-energy electron recoil events reported by XENON1T, many new physics scenarios have been proposed as a possible origin of the excess. One possible explanation is that the excess is a result of a fast moving dark matter (DM), with velocity $v\sim0.05-0.20$ and mass between 1 MeV and 10 GeV, scattering off an electron. Assuming the fast moving DM-electron interaction is mediated by a vector particle, we derive collider constraints on the said DM-electron interaction. The bounds on DM-electron coupling is then used to constrain possible production mechanisms of the fast moving DM. We find that the preferred mass of the vector mediator is relatively light ($\lesssim$ 1 GeV) and the coupling of the vector to the electron is much smaller than the coupling to the fast moving DM.
\end{abstract}

\maketitle

\section{Introduction} \label{sec:intro}
%
Recently the XENON Collaboration has reported excess of electron recoil events in the data collected during the Science Run 1 of the XENON1T detectors~\cite{Aprile:2020tmw}. At the moment, we do not know the precise origin of the excess. The excess could be a hint of new physics or it could turn out to be an unaccounted background in the detector. Nevertheless, it is interesting to interpret the result in terms of new physics. 

The XENON Collaboration gives three new physics interpretations of the electron recoil excess: 1) the absorption of the solar axion~\cite{vanBibber:1988ge,Moriyama:1995bz,Redondo:2013wwa}, 2) the scattering of neutrino with a large magnetic dipole moment~\cite{Bell:2005kz,Bell:2006wi} and 3) the absorption of a light bosonic dark matter~\cite{Arias:2012az,An:2014twa}. Of the three new physics scenarios, only the solar axion and the neutrino scattering of the excess result in a statistical significance greater than 3 sigma. Interestingly, taken at face value the preferred parameter space for both explanations seem to be ruled out by stellar cooling constraints~\cite{Aprile:2020tmw}. Moreover, it has been pointed out that the solar axion interpretation is also ruled out by  astrophysical observations~\cite{luzio2020solar}. In fact, Ref.~\cite{buch2020galactic} shows that even a broader interpretation of the  excess in terms of an absorption of an axion (or any relativistic boson) from the galactic center is incompatible with existing constraints. 
On the other hand, constraints on the neutrino scattering interpretation can be relaxed if one introduces a new particle mediating the interaction with the electron~\cite{bally2020neutrino}.
For the bosonic dark matter absorption interpretation, even though the statistical significance of the excess is low, it should not be discounted. See Refs.~\cite{Takahashi:2020bpq,Alonso-Alvarez:2020cdv,choi2020xenon1t} for interesting models employing this scenario. 

In addition to the above interpretations, the XENON1T excess can also be explained by a flux of fast moving dark matter~\cite{Kannike:2020agf,Fornal:2020npv,chen2020sun,Du:2020ybt,Su:2020zny} with velocity $v\sim~0.05-0.20$ and mass $m_{DM}~1$MeV - 10 GeV. In order to explain the XENON1T excess in this scenario, the product of dark matter density ($n_{DM}$) times the dark matter electron scattering cross-section at a momentum exchange equal to an inverse Bohr's radius ($\sigma_e$) has to be of order $10^{-44}-10^{-43}$ cm$^{-1}$~\cite{Kannike:2020agf}. 

In this work we derive constraints on such a class of dark matter models where a fast moving component would have to satisfy in order to successfully explain the XENON1T excess. 
In order for our constraints to be as model independent as possible we will work in a context of a simplified model. 
This allows us to utilize collider observables as well as astrophysical measurements to constrain parameter space of the model. These constraints help us better identify a possible source of the fast moving dark matter component.

It should be noted that scenarios for explaining the XENON1T excess mentioned above are by no means exhaustive, see Refs.~\cite{Harigaya:2020ckz,paz2020shining,bell2020explaining,Boehm:2020ltd,sierra2020light} for other possible interpretations of the electron recoil excess.




The paper is structured as follows. In Sec.~\ref{sec:model} we  set up our simplified model for the XENON1T electron recoil excess. 
In Sec.~\ref{sec:constraints} we discuss relevant constraints from both collider experiments and astrophysical observations.
We derive constraints on our simplified model for a specific set of benchmark scenarios in Sec~\ref{sec:result}. Finally, we conclude in Sec.~\ref{sec:conclusion}.

\section{Models for XENON1T Excess} \label{sec:model}
We consider a scenario where the fast-moving DM $\psi$ couples to electron via a vector boson mediator $Z'$. 
For simplicity, we will assume $\psi$ is a fermion.
The $\psi$ could be the only particle in the dark sector. In this case, the flux of fast moving DM arises from a semi-annihilation process $\psi+\psi\to\bar\psi+X$~\cite{DEramo:2010keq,Kamada:2017gfc,Smirnov:2020zwf},
where $X$ can be any particle that is neutral under a symmetry that stabilizes $\psi$. Assuming that most of the $\psi$ particles are cold DM, the velocity of $\psi$ after the semi-annihilation process will be boosted by
\begin{equation}
	\gamma_\psi = \frac{5m_\psi^2-m_X^2}{4m_\psi^2}.
\end{equation}
In order for $\psi$ to achieve velocity $v\sim0.05-0.20$ (to explain the XENON1T excess), one needs $m_X/m_\psi\sim0.956-0.997$.

It is also possible that $\psi$ is part of a more complicated dark sector. For simplicity, we assume a two-component DM scenario, i.e., $\chi$ and $\psi$. Here, we take $\chi$ to be cold DM. The nature of the DM $\chi$ is not important for the present analysis; it can be a scalar or a fermion. The flux of a fast-moving $\psi$ could arise from an annihilation process $\chi\bar\chi\to\psi\bar\psi$ or the decay $\chi\to\psi\bar\psi$. In the annihilation case, one needs a nearly degenerate $m_\chi$ and $m_\psi$ to achieve the fast velocity $v_\psi\sim0.05-0.20$. On the other hand, in the decay case, one needs $m_\chi\simeq 2m_\psi$ to achieve the desired velocity.
 
In this work, we will not be concerned about the specifics of the model. Instead, we will only focus on the interactions between $\psi$ and electron mediated by a  $Z'$.  The interaction terms relevant for our analysis are
\begin{equation}
	\mathcal{L}_{Z'} \supset -ig_\psi\bar\psi\gamma^\mu\psi Z'_\mu -ig_e\bar e\gamma^\mu eZ'_\mu.
	\label{eq:lag}
\end{equation}
We will parametrize the $\psi$ flux in terms of the annihilation cross-section $\langle\sigma v\rangle$ in the case of $\psi$ being produced from DM (semi-)annihilation. If $\psi$ is produced by a decay of $\chi$, the flux can be parametrized in terms of the partial decay-width $\Gamma$. We also allow for the possibility that the flux of $\psi$ is originated from the annihilation/decay of DM captured in the Sun. Assuming the DM capture rate and the depletion rate are in equilibrium, the flux of $\psi$ can be related to the DM-proton cross section $\sigma_{\chi p}$~\cite{Hooper:2018kfv}.
By parametrizing the flux in terms of $\langle\sigma v\rangle$/$\Gamma$/$\sigma_{\chi p}$, our results obtained by the analysis below will be applicable to any $\psi$ production channels. 

In order to explain the XENON1T excess, we need $n_{\psi}\sigma_e\sim10^{-44}-10^{-43}$ cm$^{-1}$~\cite{Kannike:2020agf}. In terms of our model parameter we have $n_\psi$ is proportional to $\langle\sigma v\rangle$, $\Gamma$ or $\sigma_{\chi p}$, while $\sigma_e$ is given by~\cite{Roberts:2019chv}
\begin{equation}
	\sigma_e = \frac{a_0^2}{\pi}\frac{\alpha^2g_e^2g_\psi^2}{[(m_{Z'}/m_e)^2+\alpha^2]^2}
	\approx \frac{g_e^2g_\psi^2}{\pi}\frac{m_e^2}{m_{Z'}^4},
	\label{eq:sigmae}
\end{equation}
where $a_0 = 1/(\alpha m_e)$ is the Bohr radius, $\alpha$ is the QED fine structure constant, $m_e$ is the mass of the electron. The last approximation is valid as long as $m_{Z'}/m_e\gg \alpha$. The couplings $g_e$ and $g_\psi$ can be probed at collider experiments, discussed in Sec.~\ref{sec:collider}. Here, we quickly note that collider constraints lead to an upper  bound on $\sigma_e$. This allows us to derive the lower bounds on $\langle\sigma v\rangle$/$\Gamma$/$\sigma_{\chi p}$, depending on the source of the $\psi$ flux.

\section{Constraints} \label{sec:constraints}

\subsection{Constraints from collider physics} \label{sec:collider}
The coupling between $Z'$ and the electron leads to a production of $Z'$ at fixed-target experiments as well as collider experiments.  
Once produced, the $Z'$ can decay visibly into $e^+e^-$ or invisibly into $\psi\bar\psi$. These decay channels lead to  rich phenomenologies. For a fixed-target experiment, the E141 experiment has searched for $eN\to eNZ'; Z' \to e^+e^-$ over the 1 MeV $\lesssim m_{Z'}\lesssim 15$ MeV mass range~\cite{PhysRevLett.59.755}. 
The same process has also been searched for by the NA64 experiment, which covered a slightly wider $Z'$ mass range, i.e., 1 MeV $\lesssim m_{Z'}\lesssim 24$ MeV~\cite{Banerjee:2019hmi}. The two searches yielded complimentary exclusion bounds. In addition to the search for $Z'\to e^+e^-$, NA64 also searched for $Z'$ decay invisibly over the same mass range~\cite{NA64:2019imj}.

Heavier $Z'$ mass can be probed directly at the $e^+e^-$ colliders through $e^+e^-\to \gamma Z'; Z' \to e^+e^-/\psi\bar\psi$ processes. Both processes have been searched for by the Babar Collaboration for 20 MeV $\lesssim m_{Z'}\lesssim 10$ GeV~\cite{Lees:2014xha} and 1 MeV $\lesssim m_{Z'}\lesssim 8$ GeV~\cite{Lees:2017lec}, respectively. 
For $m_{Z'}\gtrsim 8$ GeV, it can be probed by the LEP and the LHC where there exist several searches relevant for our model. If the $Z'$ decays dominantly into $e^+e^-$, we find that the most constraining bounds for $m_{Z'}\lesssim 70$ GeV come from the partial decay width of $Z$ to 4 leptons measurement~\cite{PhysRevD.98.030001}. 
For $m_{Z'} \gtrsim $ 70 GeV, we find that the LEP searches for neutralino R-parity-violating (RPV) decays~\cite{Barate:1999qx,Alcaraz:2006mx} places the strongest bound on our model. In deriving the LEP RPV bounds, we follow the procedure outlined in Ref.~\cite{Freitas:2014pua}. Finally, if $Z'$ decays dominantly into invisible, we find that the LEP monophoton searches~\cite{Abdallah:2008aa} provide the strongest bounds. Here, we follow the procedure outlined in Ref.~\cite{Fox:2011fx} to obtain the LEP monophoton bounds.  

    
In addition to collider bounds, the coupling of $Z'$ to electron is also constrained by the anomalous magnetic dipole moment of the electron measurement. In our model, such dipole moment constraint is relevant for the mass range 10 MeV $\lesssim m_{Z'}\lesssim$ 40 MeV. 
    
In order to avoid having all these different constraints being cluttered in our final results, we will group the bounds together as follow. We collectively refer to the E141, the NA64 and the Babar searches for $\gamma e^+e^-$ signature in the final state, together with constraints from electron anomalous magnetic dipole moment, as the low-energy bound. Similarly, the constraints from NA64 and Babar searches for $\gamma+$invisible final states will be referred to as the low-energy monophoton search. 

With all the constraints mentioned above, we can derive an upper bound on $\sigma_e$, see Eq.~\eqref{eq:sigmae}, for any set of $m_\psi$, $m_{Z'}$, $g_e$ and $g_\psi$. We then use this upper bound on $\sigma_e$ to deduce the minimum flux of $\psi$ required to produce XENON1T electron recoil excess. Finally, we interpret the minimum flux in terms of the lower bound on $\langle\sigma v\rangle$/$\Gamma$/$\sigma_{\chi p}$ depending on the mechanism responsible for generating the $\psi$ flux.

\subsection{Model Independent Constraints} \label{sec:independent}
Astrophysical observations can be used to derive model independent bounds on $\langle\sigma v\rangle$ and $\Gamma$. For $\Gamma$, the partial decay width to $\psi\bar\psi$ is robustly constrained by the age of the universe. Thus we  must have $\Gamma^{-1}\gtrsim4\times10^{17}$ s. Similarly, the annihilation cross-section $\langle\sigma v\rangle$ for $\chi\bar\chi\to\psi\bar\psi$ is constrained by the Kaplinghat, Knox and Turner (KKT) bound~\cite{Kaplinghat:2000vt,Beacom:2006tt}
\begin{equation}
	\langle\sigma v\rangle \lesssim 3\times10^{-19}\frac{\text{cm}^3}{\text{s}}\frac{m_\chi}{\rm GeV}.
\end{equation}

For the DM-proton cross-section $\sigma_{\chi p}$, DM direct detection experiments provide a robust bound for $m_\chi\gtrsim100$ MeV~\cite{Aprile:2019jmx,Amole:2019fdf}. Since in general the bound on spin-dependent cross section is weaker than the spin-independent one, we will use the more conservative spin-dependent cross-section bound as our model independent constraint. For lighter $m_\chi$, the $\sigma_{\chi p}$ is constrained by cosmic microwave background (CMB) data~\cite{Gluscevic:2017ywp,Xu:2018efh}. To be conservative, we assume the canonical spin-independent and velocity-independent $\sigma_{\chi p}$ for the CMB anisotropy measurements, which constrain $\sigma_{\chi p}\in[10^{-26},10^{-25}]$ cm$^{2}$ for 1 MeV $\lesssim m_\chi\lesssim 100$ MeV~\cite{Slatyer:2018aqg}. 

\section{Results} \label{sec:result}
The analysis in Ref.~\cite{Kannike:2020agf} suggested 1 MeV $\lesssim m_\psi\lesssim 10$ GeV with $0.05\lesssim v\lesssim 0.20$ can give a good fit to the XENON1T excess. However, the analysis of Ref.~\cite{Fornal:2020npv} seems to favor only $m_\psi=10$ GeV with $v=0.06$. Thus in our numerical analysis, we will take as our benchmark cases $m_\psi=1$ MeV and $10$ GeV. To further reduce the model parameters, we consider two different cases for the coupling $g_\psi$ and $g_e$. In the first case, we fix $g_\psi=1$ and treat $g_e$ as a free parameter, while in the second case we take $g_\psi=g_e$. 

For each of our benchmark scenario ($m_\psi$, $g_\psi$) = $\{$(1 MeV, 1), (1 MeV, $g_e$), (10 GeV, 1), (10 GeV, $g_e$)$\}$, we derive the collider bounds on the couplings as a function of $m_{Z'}$. We then interpret these bounds in terms of lower bounds on the annihilation cross-section ($\langle\sigma v\rangle$), the partial decay width ($\Gamma$) and the nucleon cross-section ($\sigma_{\chi N}$) require to fit the XENON1T electron recoil excess. We also compare our collider bounds against other relevant model independent bounds discussed in Sec.~\ref{sec:independent}. 

\begin{figure}[h!]
     \centering
        \includegraphics[width=0.4\textwidth]{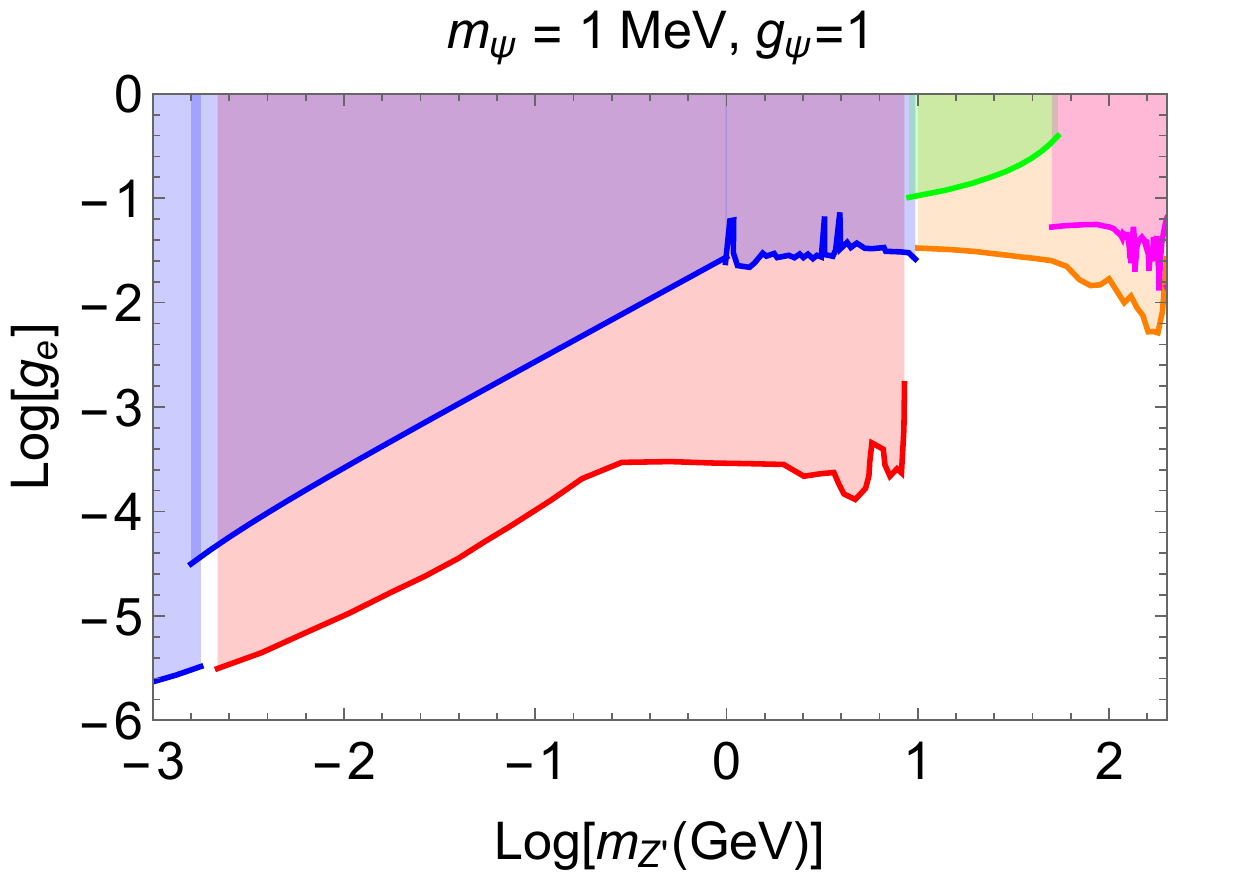}
        \includegraphics[width=0.4\textwidth]{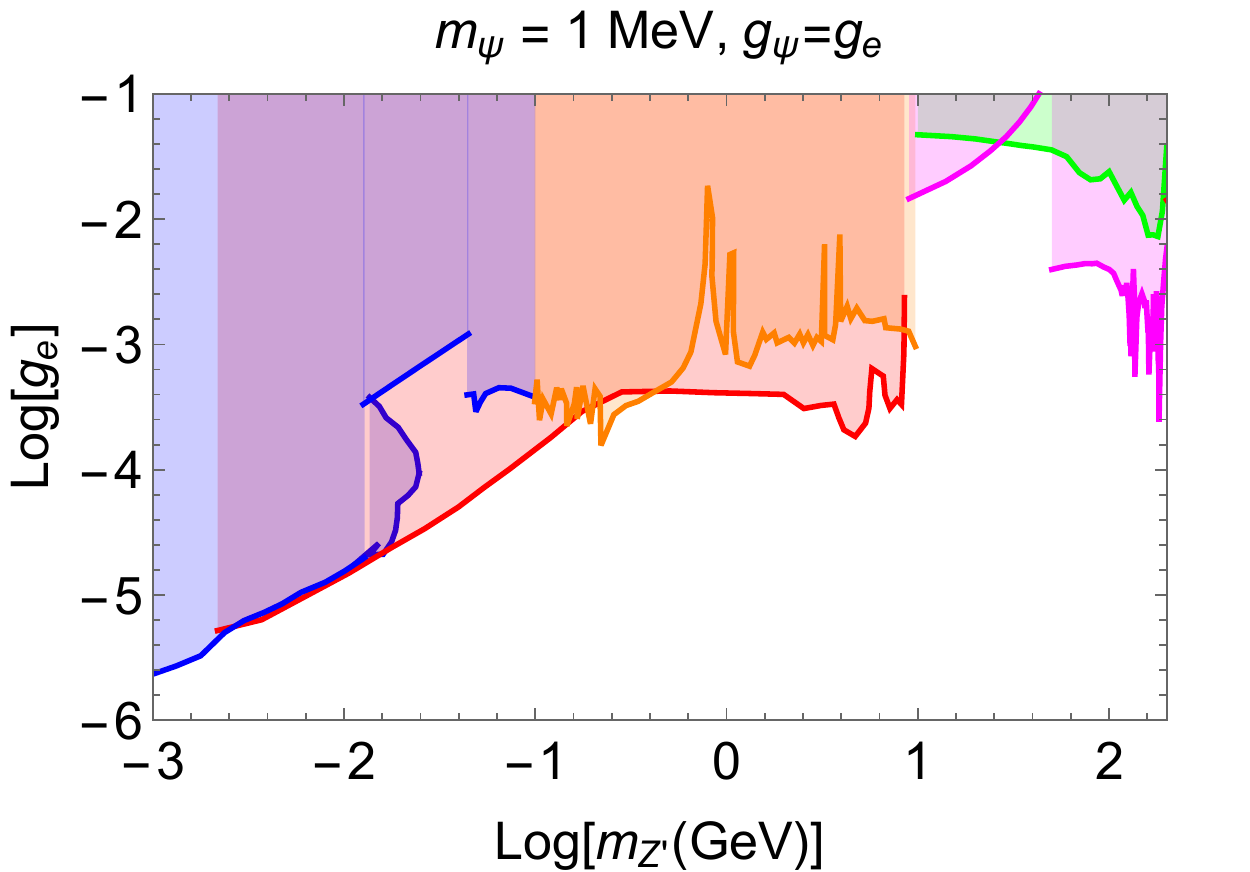}
        \caption{Collider bounds on the coupling $g_e$ as a function of $m_{Z'}$ with $m_\psi=1$ MeV. The constraints from low-energy bound, low-energy monophoton search, $Z\to4\ell$, LEP RPV and LEP monophoton searches are shown in blue, red, green, magenta and orange respectively.}
        \label{fig:gbound}
\end{figure}
We first consider a light DM scenario $m_\psi=1$ MeV. The collider bounds on the coupling $g_e$ are shown in Fig.~\ref{fig:gbound}. We find that for the case $g_\psi=1$, the coupling $g_e$ is constrained so that $g_e\ll g_\psi$ especially for $m_{Z'}\lesssim$ 10 GeV.~\footnote{The hierarchy between $g_\psi$ and $g_e$ can be easily accommodated in a model where $g_e$ is induced by kinetic mixing of the $Z'$ and the photon.} Note that in this case, the monophoton bounds are especially strong because $Z'$ decays dominantly into invisible. In the case $g_\psi=g_e$, we find that both couplings have to be small with $g_\psi=g_e\ll1$. However, in this case the low-energy monophoton bounds and the low-energy bounds are comparable because the decay rate $Z'\to e^+e^-$ and $Z'\to$invisible are approximately the same.  

\begin{figure}[h!]
     \centering
        \includegraphics[width=0.4\textwidth]{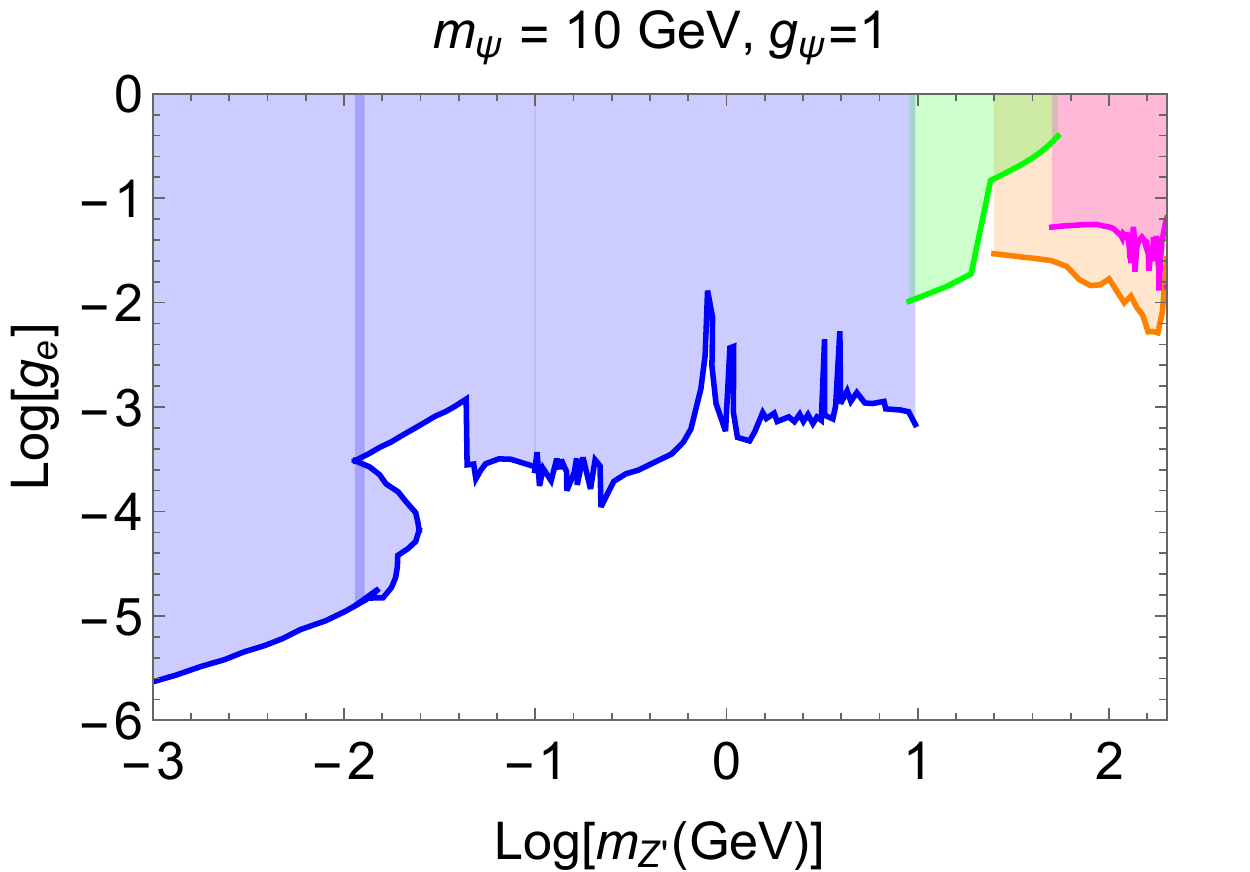}
        \includegraphics[width=0.4\textwidth]{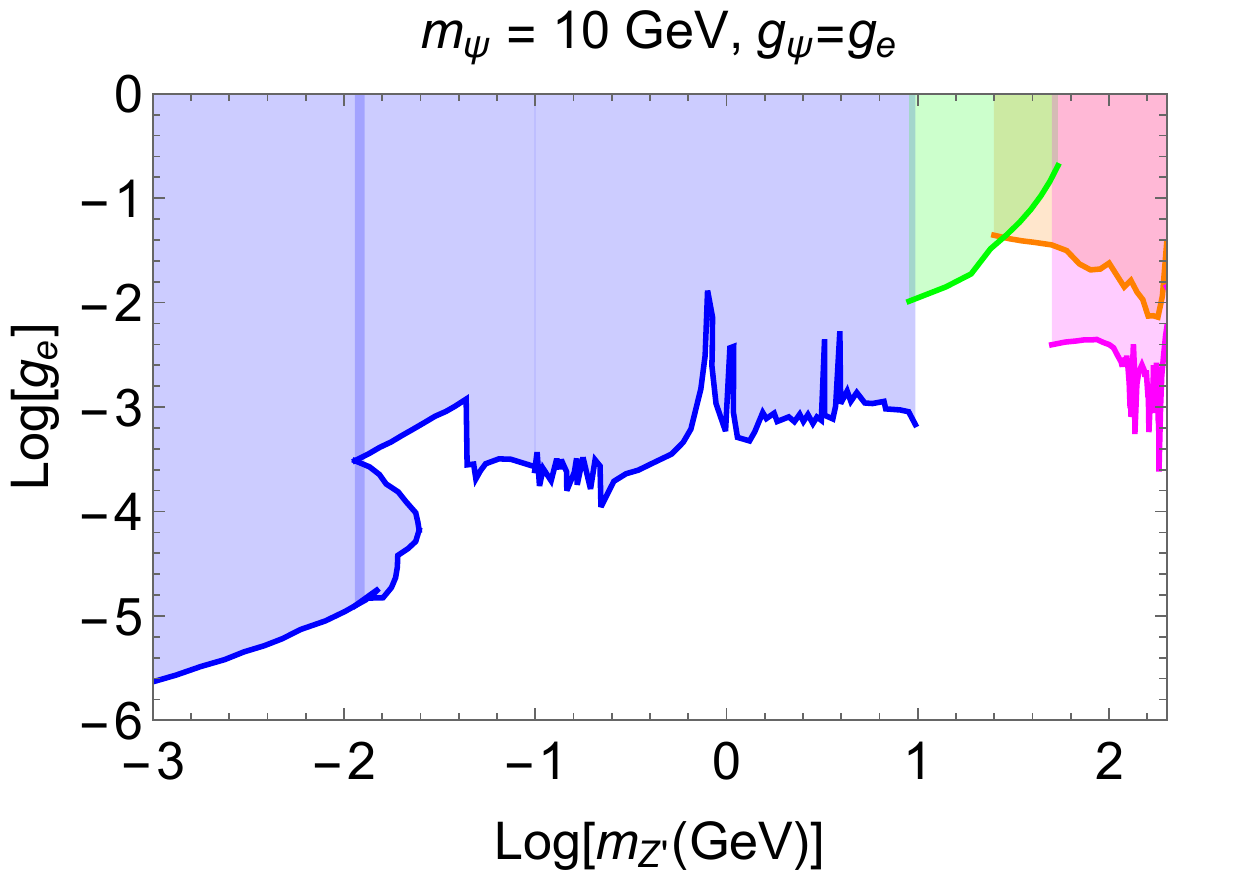}
        \caption{Collider bounds on the coupling $g_e$ as a function of $m_{Z'}$ with $m_\psi=10$ GeV. The constraints from low-energy bound, low-energy monophoton search, $Z\to4\ell$, LEP RPV and LEP monophoton searches are shown in blue, red, green, magenta and orange respectively.}
        \label{fig:gbound2}
\end{figure}
For the case of $m_\psi=10$ GeV, the collider constraints on the coupling $g_e$ are shown in Fig.~\ref{fig:gbound2}. Similar to the case where $m_\psi=1$ MeV, the coupling $g_e$ in this case is also constrained such that $g_e\ll1$. The main difference from the $m_\psi=1$ MeV case is that for $m_{Z'}\lesssim10$ GeV there are no low-energy monophoton bounds. As a result the low-energy bounds become more constraining.

\begin{figure}[h!]
     \centering
         \includegraphics[width=0.4\textwidth]{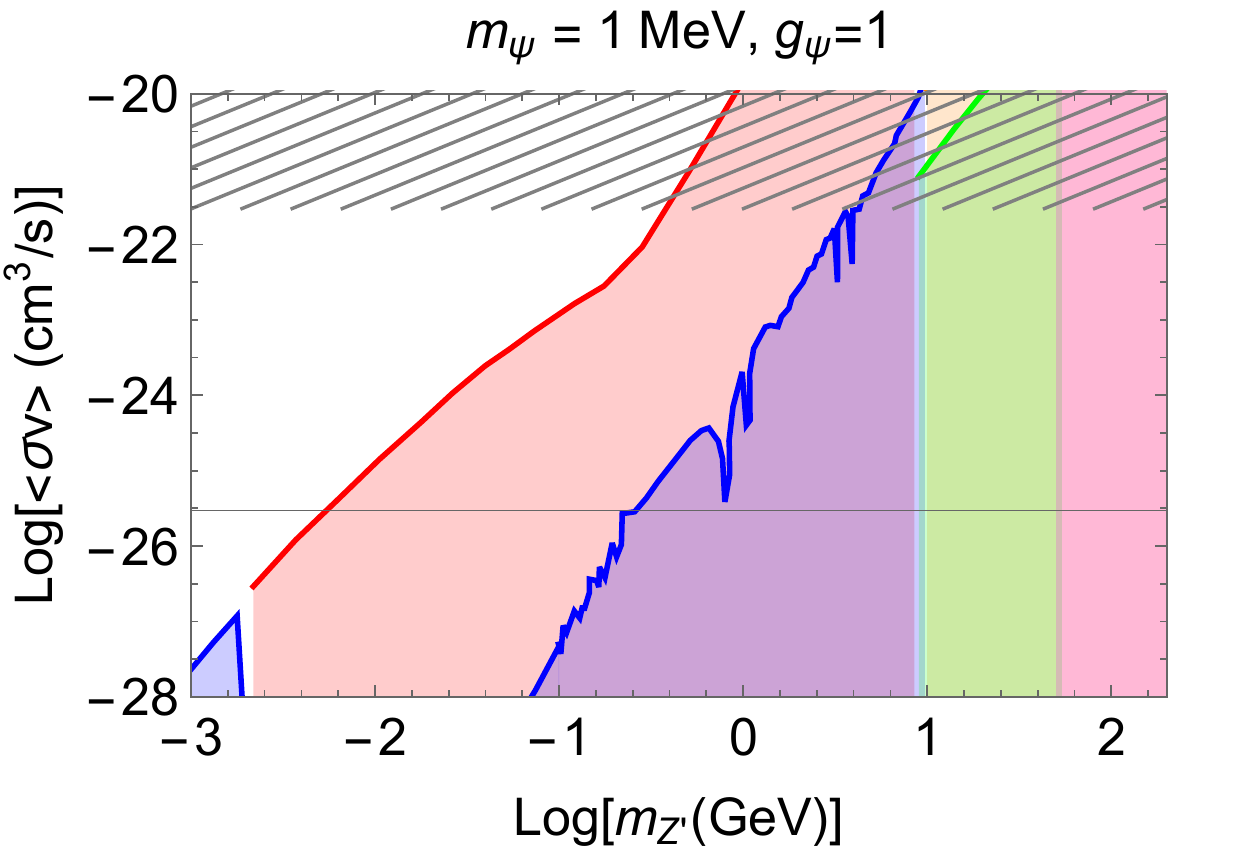}
         \includegraphics[width=0.4\textwidth]{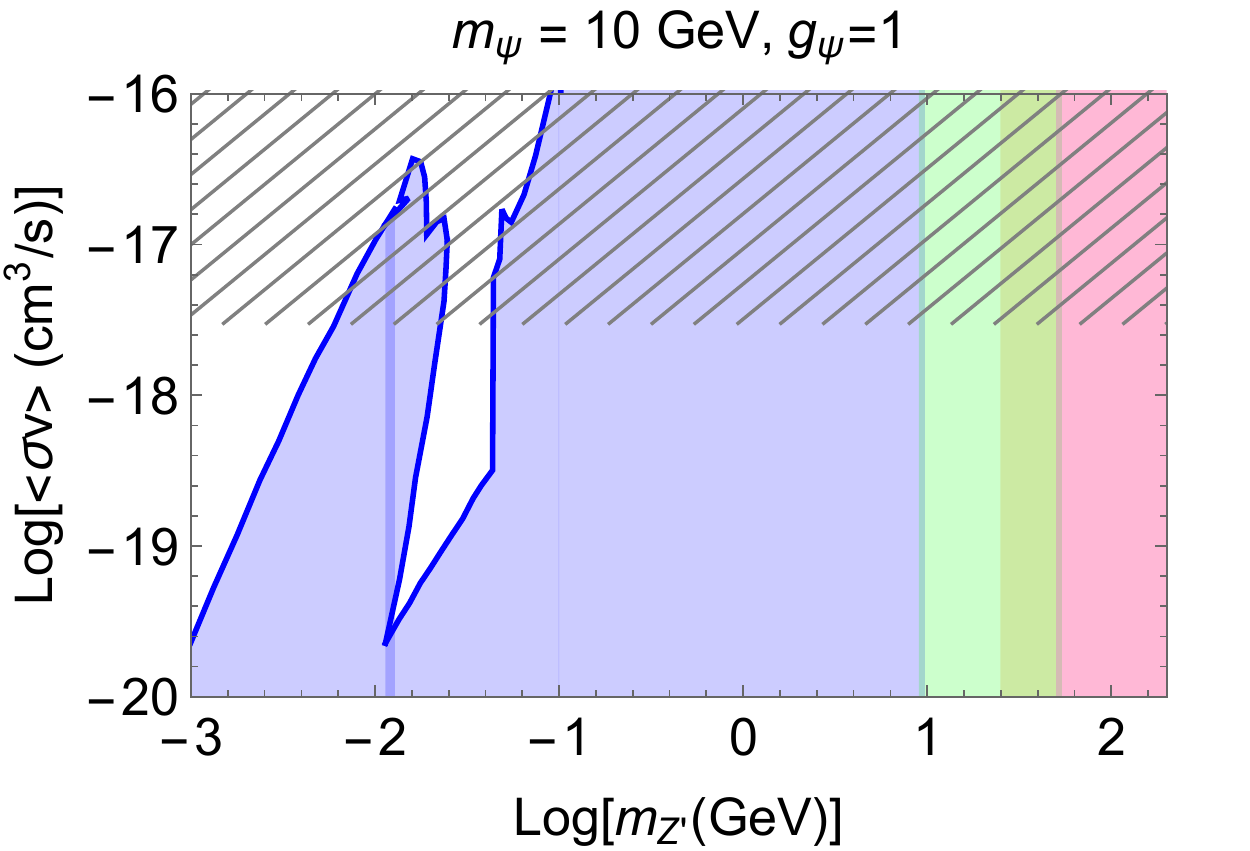}
        \caption{The lower bound on the annihilation cross-section as a function of $m_{Z'}$ with $g_\psi=1$. The constraints from low-energy bound, low-energy monophoton search, $Z\to4\ell$, LEP RPV and LEP monophoton searches are shown in blue, red, green, magenta and orange respectively. The grey hashed region is excluded by the KKT bound. The horizontal line in the left figure represents the thermal relic cross-section, $\langle\sigma v\rangle\approx3\times10^{-26}$ cm$^3$/s.}
        \label{fig:sigmav}
\end{figure}
We turn next to analyze the bound on the $\psi$ production mechanisms. In the case that $\psi$ is produced from DM annihilation, the bounds on $\langle\sigma v\rangle$ are shown in Fig.~\ref{fig:sigmav}. Of all our benchmark cases, only the $g_\psi=1$ scenarios are compatible with both the collider constraints and the model independent constraints. Note that the mass of the $Z'$ is tightly constrained for both $m_\psi=1$ MeV and 10 GeV benchmark case. We find that $m_{Z'}\lesssim300$ MeV for $m_\psi=1$ MeV and case $m_{Z'}\lesssim30$ MeV for $m_\psi=10$ GeV.

\begin{figure}[h!]
     \centering
         \includegraphics[width=0.4\textwidth]{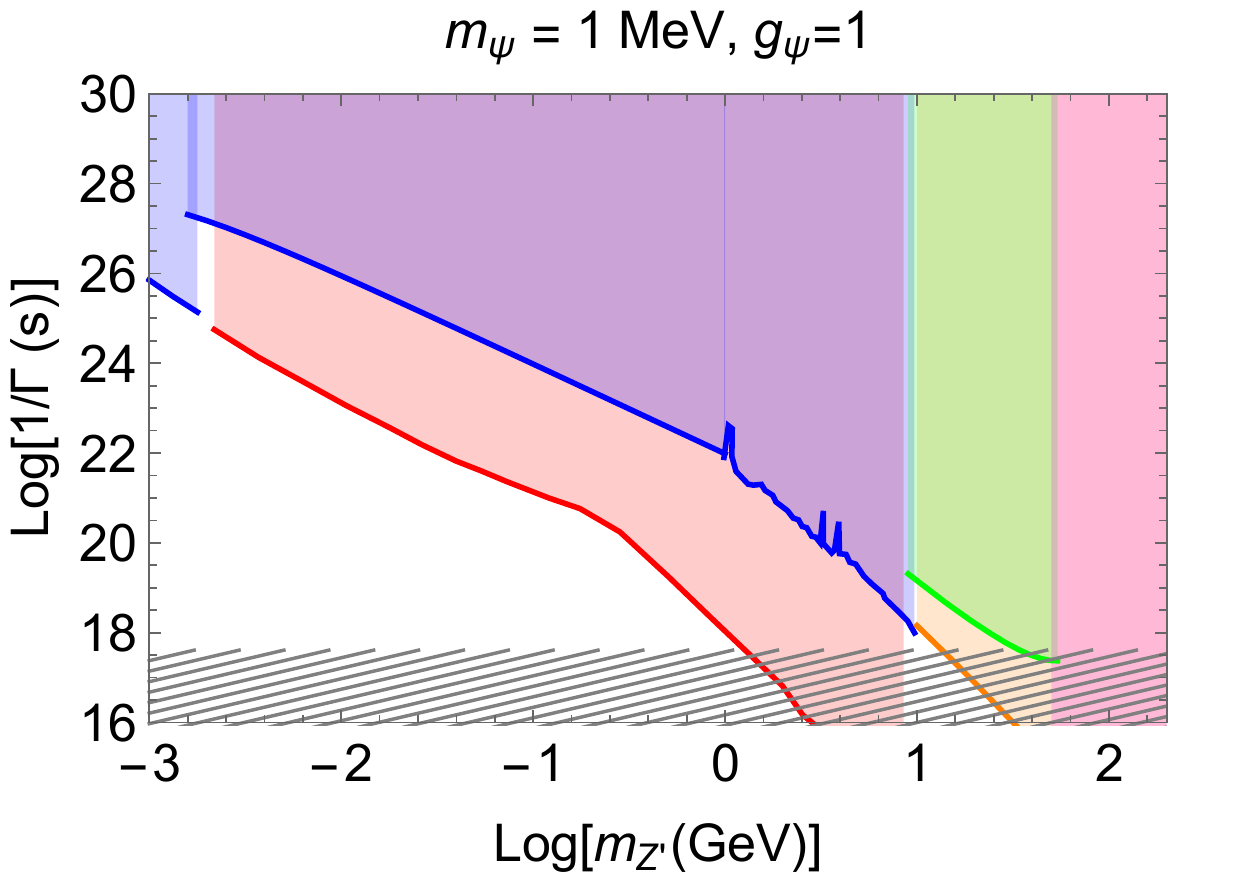}
         \includegraphics[width=0.4\textwidth]{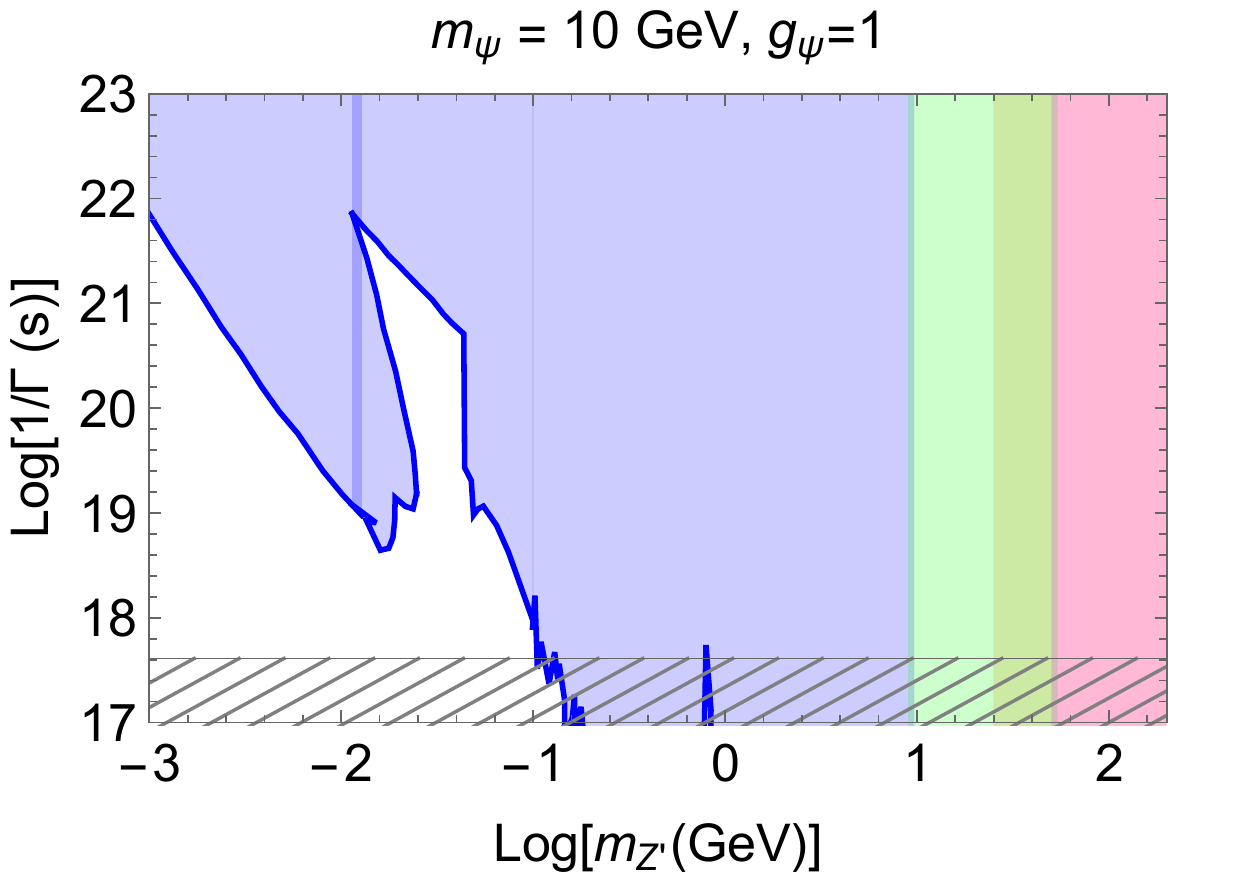}
        \caption{The upper bound on the inverse partial decay as a function of $m_{Z'}$ with $g_\psi=1$. The constraints from low-energy bound, low-energy monophoton search, $Z\to4\ell$, LEP RPV and LEP monophoton searches are shown in blue, red, green, magenta and orange respectively. In the grey hashed region the $\chi$ life time is shorter than the age of the universe.}
        \label{fig:Gamma}
\end{figure}
Next we consider the $\psi$ production via the decay of $\chi$. Similar to the annihilation production case, only the $g_\psi=1$ scenarios are compatible with all the constraints, see Fig.~\ref{fig:Gamma}. However, in this case the constraints on the $Z'$ mass is slightly relaxed. For $m_\psi=1$ MeV, the mass of the $Z'$ is constrained to be $m_{Z'}\lesssim1$ GeV. For a heavier DM mass, $m_\psi=10$ GeV, we find that $m_{Z'}\lesssim100$ MeV.

\begin{figure}[h!]
     \centering
         \includegraphics[width=0.4\textwidth]{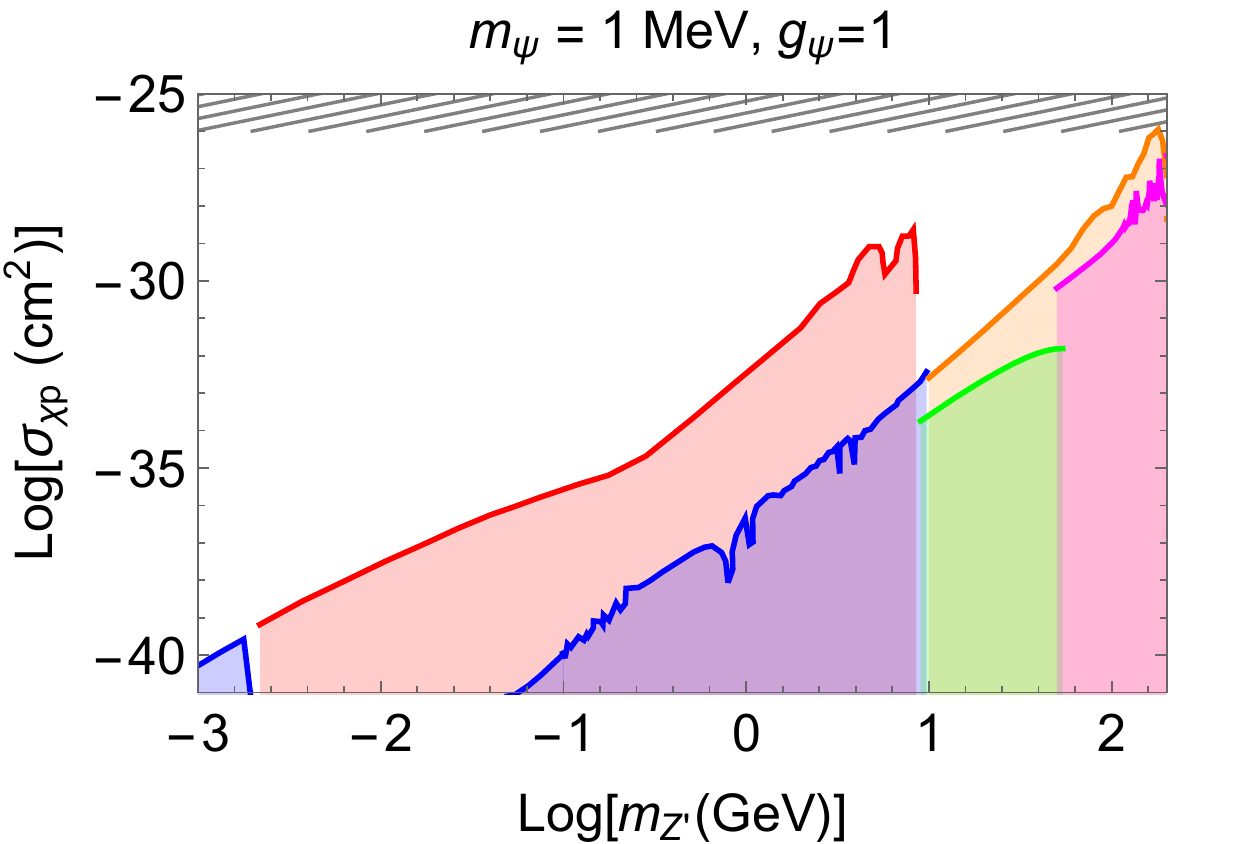}
         \includegraphics[width=0.4\textwidth]{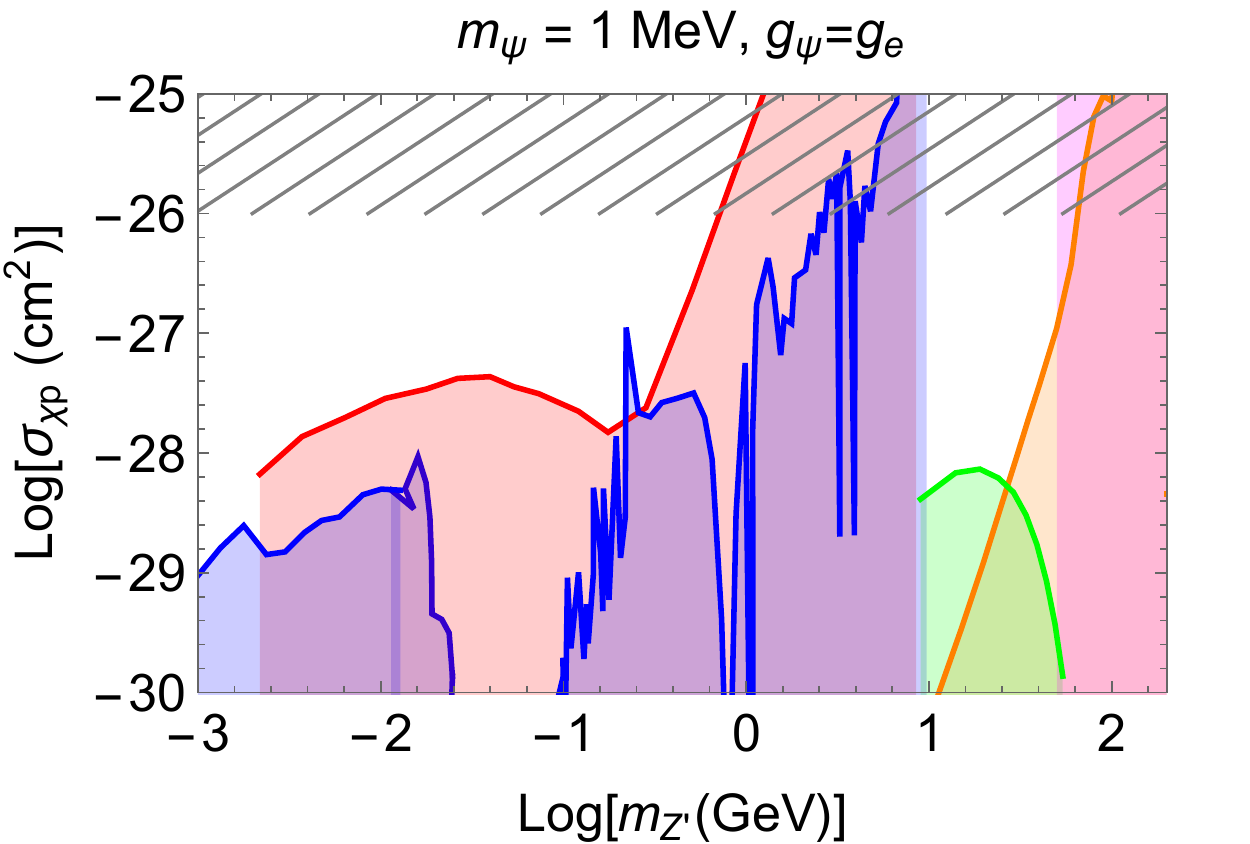}
        \caption{The lower bound on the DM-proton as a function of $m_{Z'}$ with $m_\psi=1$ MeV. The constraints from low-energy bound, low-energy monophoton search, $Z\to4\ell$, LEP RPV and LEP monophoton searches are shown in blue, red, green, magenta and orange respectively. The grey hashed region is excluded by the CMB anisotropy measurements.}
        \label{fig:proton}
\end{figure}
Last, but not least, we consider the $\psi$ production in the Sun. 
In this scenario, the benchmark cases $m_\psi=10$ GeV are severely constrained by the DM direct detection experiments such that the region of parameter space accommodating the XENON1T excess has been excluded. On the other hand, the benchmark cases $m_\psi=1$ MeV fall below the threshold for direct detection experiment. As a result, they are only subjected to milder constraints from CMB anisotropy measurements, see Fig.~\ref{fig:proton}. One can see that this production scenario is the least constrained one. In the case $g_\psi=1$, the $Z'$ mass is loosely constrained, $m_{Z'}\lesssim200$ GeV. In the case $g_\psi=g_e$ the $Z'$ mass is constrained to lie in the range $m_{Z'}\lesssim500$ MeV or 10 GeV $\lesssim m_{Z'}\lesssim50$ GeV.

\section{Conclusions and Discussions} \label{sec:conclusion}

In this work, we study collider constraints on the class of models that explain the XENON1T excess by having a fast moving DM scattered off an electron. The bounds obtained from our analysis should be applicable to any models in which the fast DM interact with the electron via a vector exchange.  We derive the collider bounds on the vector-electron coupling in Figs.~\ref{fig:gbound} and~\ref{fig:gbound2}. 
We also interpret our bounds in terms of the production mechanisms of the fast moving DM: annihilation of heavier DM inside galaxy, decay of heavier DM decay inside the galaxy, and annihilation/decay of heavier DM captured in the Sun. 

In all the benchmark scenarios considered here, we find that a light fast moving DM is less constraining than a heavy fast moving DM in explaining the excess in the XENON1T electron recoil events, see e.g., Figs.~\ref{fig:sigmav} and ~\ref{fig:Gamma}. We also find that the mass of the vector mediator must be lighter than $1$ GeV except for the case where the fast moving DM is produced from DM captured in the Sun, see Fig.~\ref{fig:proton}.  

One might argue that our analysis is too simplified because we only consider two decay channels for the $Z'$: $Z'\to e^+e^-$ and $Z\to\psi\bar\psi$. The latter is invisible. We quickly comment on the effect of considering additional $Z'$ decay channels. If the additional decay mode is invisible, it will result in a stronger collider bounds, thanks to the monophoton analysis. If the additional decay channel is visible, it will dilute the bounds considered in our analysis. However, it will also lead to a richer collider signatures~\cite{Essig:2013lka} that, in principle, can be included into our analysis. Thus we do not expect our collider bounds to change significantly.  

\bigskip
\paragraph*{Acknowledgments:}
%
The work of RP is supported by the Parahyangan Catholic University under grant no.
III/LPPM/2020-01/02-P.
The work of PU has been supported in part by the Thailand Research Fund under contract no.~MRG6280186, the Faculty of Science, Srinakharinwirot University under grant no.~222/2562 and 497/2562, and by the National Astronomical Research Institute of Thailand.

\appendix

\bibliography{ref} \bibliographystyle{apsrev4-1}

\begin{thebibliography}{48}%
\makeatletter
\providecommand \@ifxundefined [1]{%
 \@ifx{#1\undefined}
}%
\providecommand \@ifnum [1]{%
 \ifnum #1\expandafter \@firstoftwo
 \else \expandafter \@secondoftwo
 \fi
}%
\providecommand \@ifx [1]{%
 \ifx #1\expandafter \@firstoftwo
 \else \expandafter \@secondoftwo
 \fi
}%
\providecommand \natexlab [1]{#1}%
\providecommand \enquote  [1]{``#1''}%
\providecommand \bibnamefont  [1]{#1}%
\providecommand \bibfnamefont [1]{#1}%
\providecommand \citenamefont [1]{#1}%
\providecommand \href@noop [0]{\@secondoftwo}%
\providecommand \href [0]{\begingroup \@sanitize@url \@href}%
\providecommand \@href[1]{\@@startlink{#1}\@@href}%
\providecommand \@@href[1]{\endgroup#1\@@endlink}%
\providecommand \@sanitize@url [0]{\catcode `\\12\catcode `\$12\catcode
  `\&12\catcode `\#12\catcode `\^12\catcode `\_12\catcode `\%12\relax}%
\providecommand \@@startlink[1]{}%
\providecommand \@@endlink[0]{}%
\providecommand \url  [0]{\begingroup\@sanitize@url \@url }%
\providecommand \@url [1]{\endgroup\@href {#1}{\urlprefix }}%
\providecommand \urlprefix  [0]{URL }%
\providecommand \Eprint [0]{\href }%
\providecommand \doibase [0]{http://dx.doi.org/}%
\providecommand \selectlanguage [0]{\@gobble}%
\providecommand \bibinfo  [0]{\@secondoftwo}%
\providecommand \bibfield  [0]{\@secondoftwo}%
\providecommand \translation [1]{[#1]}%
\providecommand \BibitemOpen [0]{}%
\providecommand \bibitemStop [0]{}%
\providecommand \bibitemNoStop [0]{.\EOS\space}%
\providecommand \EOS [0]{\spacefactor3000\relax}%
\providecommand \BibitemShut  [1]{\csname bibitem#1\endcsname}%
\let\auto@bib@innerbib\@empty
\bibitem [{\citenamefont {Aprile}\ \emph {et~al.}(2020)\citenamefont {Aprile}
  \emph {et~al.}}]{Aprile:2020tmw}%
  \BibitemOpen
  \bibfield  {author} {\bibinfo {author} {\bibfnamefont {E.}~\bibnamefont
  {Aprile}} \emph {et~al.} (\bibinfo {collaboration} {XENON}),\ }\href@noop {}
  {\  (\bibinfo {year} {2020})},\ \Eprint {http://arxiv.org/abs/2006.09721}
  {arXiv:2006.09721 [hep-ex]} \BibitemShut {NoStop}%
\bibitem [{\citenamefont {van Bibber}\ \emph {et~al.}(1989)\citenamefont {van
  Bibber}, \citenamefont {McIntyre}, \citenamefont {Morris},\ and\
  \citenamefont {Raffelt}}]{vanBibber:1988ge}%
  \BibitemOpen
  \bibfield  {author} {\bibinfo {author} {\bibfnamefont {K.}~\bibnamefont {van
  Bibber}}, \bibinfo {author} {\bibfnamefont {P.}~\bibnamefont {McIntyre}},
  \bibinfo {author} {\bibfnamefont {D.}~\bibnamefont {Morris}}, \ and\ \bibinfo
  {author} {\bibfnamefont {G.}~\bibnamefont {Raffelt}},\ }\href {\doibase
  10.1103/PhysRevD.39.2089} {\bibfield  {journal} {\bibinfo  {journal} {Phys.
  Rev. D}\ }\textbf {\bibinfo {volume} {39}},\ \bibinfo {pages} {2089}
  (\bibinfo {year} {1989})}\BibitemShut {NoStop}%
\bibitem [{\citenamefont {Moriyama}(1995)}]{Moriyama:1995bz}%
  \BibitemOpen
  \bibfield  {author} {\bibinfo {author} {\bibfnamefont {S.}~\bibnamefont
  {Moriyama}},\ }\href {\doibase 10.1103/PhysRevLett.75.3222} {\bibfield
  {journal} {\bibinfo  {journal} {Phys. Rev. Lett.}\ }\textbf {\bibinfo
  {volume} {75}},\ \bibinfo {pages} {3222} (\bibinfo {year} {1995})},\ \Eprint
  {http://arxiv.org/abs/hep-ph/9504318} {arXiv:hep-ph/9504318} \BibitemShut
  {NoStop}%
\bibitem [{\citenamefont {Redondo}(2013)}]{Redondo:2013wwa}%
  \BibitemOpen
  \bibfield  {author} {\bibinfo {author} {\bibfnamefont {J.}~\bibnamefont
  {Redondo}},\ }\href {\doibase 10.1088/1475-7516/2013/12/008} {\bibfield
  {journal} {\bibinfo  {journal} {JCAP}\ }\textbf {\bibinfo {volume} {12}},\
  \bibinfo {pages} {008} (\bibinfo {year} {2013})},\ \Eprint
  {http://arxiv.org/abs/1310.0823} {arXiv:1310.0823 [hep-ph]} \BibitemShut
  {NoStop}%
\bibitem [{\citenamefont {Bell}\ \emph {et~al.}(2005)\citenamefont {Bell},
  \citenamefont {Cirigliano}, \citenamefont {Ramsey-Musolf}, \citenamefont
  {Vogel},\ and\ \citenamefont {Wise}}]{Bell:2005kz}%
  \BibitemOpen
  \bibfield  {author} {\bibinfo {author} {\bibfnamefont {N.~F.}\ \bibnamefont
  {Bell}}, \bibinfo {author} {\bibfnamefont {V.}~\bibnamefont {Cirigliano}},
  \bibinfo {author} {\bibfnamefont {M.~J.}\ \bibnamefont {Ramsey-Musolf}},
  \bibinfo {author} {\bibfnamefont {P.}~\bibnamefont {Vogel}}, \ and\ \bibinfo
  {author} {\bibfnamefont {M.~B.}\ \bibnamefont {Wise}},\ }\href {\doibase
  10.1103/PhysRevLett.95.151802} {\bibfield  {journal} {\bibinfo  {journal}
  {Phys. Rev. Lett.}\ }\textbf {\bibinfo {volume} {95}},\ \bibinfo {pages}
  {151802} (\bibinfo {year} {2005})},\ \Eprint
  {http://arxiv.org/abs/hep-ph/0504134} {arXiv:hep-ph/0504134} \BibitemShut
  {NoStop}%
\bibitem [{\citenamefont {Bell}\ \emph {et~al.}(2006)\citenamefont {Bell},
  \citenamefont {Gorchtein}, \citenamefont {Ramsey-Musolf}, \citenamefont
  {Vogel},\ and\ \citenamefont {Wang}}]{Bell:2006wi}%
  \BibitemOpen
  \bibfield  {author} {\bibinfo {author} {\bibfnamefont {N.~F.}\ \bibnamefont
  {Bell}}, \bibinfo {author} {\bibfnamefont {M.}~\bibnamefont {Gorchtein}},
  \bibinfo {author} {\bibfnamefont {M.~J.}\ \bibnamefont {Ramsey-Musolf}},
  \bibinfo {author} {\bibfnamefont {P.}~\bibnamefont {Vogel}}, \ and\ \bibinfo
  {author} {\bibfnamefont {P.}~\bibnamefont {Wang}},\ }\href {\doibase
  10.1016/j.physletb.2006.09.055} {\bibfield  {journal} {\bibinfo  {journal}
  {Phys. Lett. B}\ }\textbf {\bibinfo {volume} {642}},\ \bibinfo {pages} {377}
  (\bibinfo {year} {2006})},\ \Eprint {http://arxiv.org/abs/hep-ph/0606248}
  {arXiv:hep-ph/0606248} \BibitemShut {NoStop}%
\bibitem [{\citenamefont {Arias}\ \emph {et~al.}(2012)\citenamefont {Arias},
  \citenamefont {Cadamuro}, \citenamefont {Goodsell}, \citenamefont {Jaeckel},
  \citenamefont {Redondo},\ and\ \citenamefont {Ringwald}}]{Arias:2012az}%
  \BibitemOpen
  \bibfield  {author} {\bibinfo {author} {\bibfnamefont {P.}~\bibnamefont
  {Arias}}, \bibinfo {author} {\bibfnamefont {D.}~\bibnamefont {Cadamuro}},
  \bibinfo {author} {\bibfnamefont {M.}~\bibnamefont {Goodsell}}, \bibinfo
  {author} {\bibfnamefont {J.}~\bibnamefont {Jaeckel}}, \bibinfo {author}
  {\bibfnamefont {J.}~\bibnamefont {Redondo}}, \ and\ \bibinfo {author}
  {\bibfnamefont {A.}~\bibnamefont {Ringwald}},\ }\href {\doibase
  10.1088/1475-7516/2012/06/013} {\bibfield  {journal} {\bibinfo  {journal}
  {JCAP}\ }\textbf {\bibinfo {volume} {06}},\ \bibinfo {pages} {013} (\bibinfo
  {year} {2012})},\ \Eprint {http://arxiv.org/abs/1201.5902} {arXiv:1201.5902
  [hep-ph]} \BibitemShut {NoStop}%
\bibitem [{\citenamefont {An}\ \emph {et~al.}(2015)\citenamefont {An},
  \citenamefont {Pospelov}, \citenamefont {Pradler},\ and\ \citenamefont
  {Ritz}}]{An:2014twa}%
  \BibitemOpen
  \bibfield  {author} {\bibinfo {author} {\bibfnamefont {H.}~\bibnamefont
  {An}}, \bibinfo {author} {\bibfnamefont {M.}~\bibnamefont {Pospelov}},
  \bibinfo {author} {\bibfnamefont {J.}~\bibnamefont {Pradler}}, \ and\
  \bibinfo {author} {\bibfnamefont {A.}~\bibnamefont {Ritz}},\ }\href {\doibase
  10.1016/j.physletb.2015.06.018} {\bibfield  {journal} {\bibinfo  {journal}
  {Phys. Lett. B}\ }\textbf {\bibinfo {volume} {747}},\ \bibinfo {pages} {331}
  (\bibinfo {year} {2015})},\ \Eprint {http://arxiv.org/abs/1412.8378}
  {arXiv:1412.8378 [hep-ph]} \BibitemShut {NoStop}%
\bibitem [{\citenamefont {Luzio}\ \emph {et~al.}(2020)\citenamefont {Luzio},
  \citenamefont {Fedele}, \citenamefont {Giannotti}, \citenamefont {Mescia},\
  and\ \citenamefont {Nardi}}]{luzio2020solar}%
  \BibitemOpen
  \bibfield  {author} {\bibinfo {author} {\bibfnamefont {L.~D.}\ \bibnamefont
  {Luzio}}, \bibinfo {author} {\bibfnamefont {M.}~\bibnamefont {Fedele}},
  \bibinfo {author} {\bibfnamefont {M.}~\bibnamefont {Giannotti}}, \bibinfo
  {author} {\bibfnamefont {F.}~\bibnamefont {Mescia}}, \ and\ \bibinfo {author}
  {\bibfnamefont {E.}~\bibnamefont {Nardi}},\ }\href@noop {} {\enquote
  {\bibinfo {title} {Solar axions cannot explain the xenon1t excess},}\ }
  (\bibinfo {year} {2020}),\ \Eprint {http://arxiv.org/abs/2006.12487}
  {arXiv:2006.12487 [hep-ph]} \BibitemShut {NoStop}%
\bibitem [{\citenamefont {Buch}\ \emph {et~al.}(2020)\citenamefont {Buch},
  \citenamefont {Buen-Abad}, \citenamefont {Fan},\ and\ \citenamefont
  {Leung}}]{buch2020galactic}%
  \BibitemOpen
  \bibfield  {author} {\bibinfo {author} {\bibfnamefont {J.}~\bibnamefont
  {Buch}}, \bibinfo {author} {\bibfnamefont {M.~A.}\ \bibnamefont {Buen-Abad}},
  \bibinfo {author} {\bibfnamefont {J.}~\bibnamefont {Fan}}, \ and\ \bibinfo
  {author} {\bibfnamefont {J.~S.~C.}\ \bibnamefont {Leung}},\ }\href@noop {}
  {\enquote {\bibinfo {title} {Galactic origin of relativistic bosons and
  xenon1t excess},}\ } (\bibinfo {year} {2020}),\ \Eprint
  {http://arxiv.org/abs/2006.12488} {arXiv:2006.12488 [hep-ph]} \BibitemShut
  {NoStop}%
\bibitem [{\citenamefont {Bally}\ \emph {et~al.}(2020)\citenamefont {Bally},
  \citenamefont {Jana},\ and\ \citenamefont {Trautner}}]{bally2020neutrino}%
  \BibitemOpen
  \bibfield  {author} {\bibinfo {author} {\bibfnamefont {A.}~\bibnamefont
  {Bally}}, \bibinfo {author} {\bibfnamefont {S.}~\bibnamefont {Jana}}, \ and\
  \bibinfo {author} {\bibfnamefont {A.}~\bibnamefont {Trautner}},\ }\href@noop
  {} {\enquote {\bibinfo {title} {Neutrino self-interactions and xenon1t
  electron recoil excess},}\ } (\bibinfo {year} {2020}),\ \Eprint
  {http://arxiv.org/abs/2006.11919} {arXiv:2006.11919 [hep-ph]} \BibitemShut
  {NoStop}%
\bibitem [{\citenamefont {Takahashi}\ \emph {et~al.}(2020)\citenamefont
  {Takahashi}, \citenamefont {Yamada},\ and\ \citenamefont
  {Yin}}]{Takahashi:2020bpq}%
  \BibitemOpen
  \bibfield  {author} {\bibinfo {author} {\bibfnamefont {F.}~\bibnamefont
  {Takahashi}}, \bibinfo {author} {\bibfnamefont {M.}~\bibnamefont {Yamada}}, \
  and\ \bibinfo {author} {\bibfnamefont {W.}~\bibnamefont {Yin}},\ }\href@noop
  {} {\  (\bibinfo {year} {2020})},\ \Eprint {http://arxiv.org/abs/2006.10035}
  {arXiv:2006.10035 [hep-ph]} \BibitemShut {NoStop}%
\bibitem [{\citenamefont {Alonso-Álvarez}\ \emph {et~al.}(2020)\citenamefont
  {Alonso-Álvarez}, \citenamefont {Ertas}, \citenamefont {Jaeckel},
  \citenamefont {Kahlhoefer},\ and\ \citenamefont
  {Thormaehlen}}]{Alonso-Alvarez:2020cdv}%
  \BibitemOpen
  \bibfield  {author} {\bibinfo {author} {\bibfnamefont {G.}~\bibnamefont
  {Alonso-Álvarez}}, \bibinfo {author} {\bibfnamefont {F.}~\bibnamefont
  {Ertas}}, \bibinfo {author} {\bibfnamefont {J.}~\bibnamefont {Jaeckel}},
  \bibinfo {author} {\bibfnamefont {F.}~\bibnamefont {Kahlhoefer}}, \ and\
  \bibinfo {author} {\bibfnamefont {L.}~\bibnamefont {Thormaehlen}},\
  }\href@noop {} {\  (\bibinfo {year} {2020})},\ \Eprint
  {http://arxiv.org/abs/2006.11243} {arXiv:2006.11243 [hep-ph]} \BibitemShut
  {NoStop}%
\bibitem [{\citenamefont {Choi}\ \emph {et~al.}(2020)\citenamefont {Choi},
  \citenamefont {Suzuki},\ and\ \citenamefont {Yanagida}}]{choi2020xenon1t}%
  \BibitemOpen
  \bibfield  {author} {\bibinfo {author} {\bibfnamefont {G.}~\bibnamefont
  {Choi}}, \bibinfo {author} {\bibfnamefont {M.}~\bibnamefont {Suzuki}}, \ and\
  \bibinfo {author} {\bibfnamefont {T.~T.}\ \bibnamefont {Yanagida}},\
  }\href@noop {} {\enquote {\bibinfo {title} {Xenon1t anomaly and its
  implication for decaying warm dark matter},}\ } (\bibinfo {year} {2020}),\
  \Eprint {http://arxiv.org/abs/2006.12348} {arXiv:2006.12348 [hep-ph]}
  \BibitemShut {NoStop}%
\bibitem [{\citenamefont {Kannike}\ \emph {et~al.}(2020)\citenamefont
  {Kannike}, \citenamefont {Raidal}, \citenamefont {Veermäe}, \citenamefont
  {Strumia},\ and\ \citenamefont {Teresi}}]{Kannike:2020agf}%
  \BibitemOpen
  \bibfield  {author} {\bibinfo {author} {\bibfnamefont {K.}~\bibnamefont
  {Kannike}}, \bibinfo {author} {\bibfnamefont {M.}~\bibnamefont {Raidal}},
  \bibinfo {author} {\bibfnamefont {H.}~\bibnamefont {Veermäe}}, \bibinfo
  {author} {\bibfnamefont {A.}~\bibnamefont {Strumia}}, \ and\ \bibinfo
  {author} {\bibfnamefont {D.}~\bibnamefont {Teresi}},\ }\href@noop {} {\
  (\bibinfo {year} {2020})},\ \Eprint {http://arxiv.org/abs/2006.10735}
  {arXiv:2006.10735 [hep-ph]} \BibitemShut {NoStop}%
\bibitem [{\citenamefont {Fornal}\ \emph {et~al.}(2020)\citenamefont {Fornal},
  \citenamefont {Sandick}, \citenamefont {Shu}, \citenamefont {Su},\ and\
  \citenamefont {Zhao}}]{Fornal:2020npv}%
  \BibitemOpen
  \bibfield  {author} {\bibinfo {author} {\bibfnamefont {B.}~\bibnamefont
  {Fornal}}, \bibinfo {author} {\bibfnamefont {P.}~\bibnamefont {Sandick}},
  \bibinfo {author} {\bibfnamefont {J.}~\bibnamefont {Shu}}, \bibinfo {author}
  {\bibfnamefont {M.}~\bibnamefont {Su}}, \ and\ \bibinfo {author}
  {\bibfnamefont {Y.}~\bibnamefont {Zhao}},\ }\href@noop {} {\  (\bibinfo
  {year} {2020})},\ \Eprint {http://arxiv.org/abs/2006.11264} {arXiv:2006.11264
  [hep-ph]} \BibitemShut {NoStop}%
\bibitem [{\citenamefont {Chen}\ \emph {et~al.}(2020)\citenamefont {Chen},
  \citenamefont {Shu}, \citenamefont {Xue}, \citenamefont {Yuan},\ and\
  \citenamefont {Yuan}}]{chen2020sun}%
  \BibitemOpen
  \bibfield  {author} {\bibinfo {author} {\bibfnamefont {Y.}~\bibnamefont
  {Chen}}, \bibinfo {author} {\bibfnamefont {J.}~\bibnamefont {Shu}}, \bibinfo
  {author} {\bibfnamefont {X.}~\bibnamefont {Xue}}, \bibinfo {author}
  {\bibfnamefont {G.}~\bibnamefont {Yuan}}, \ and\ \bibinfo {author}
  {\bibfnamefont {Q.}~\bibnamefont {Yuan}},\ }\href@noop {} {\enquote {\bibinfo
  {title} {Sun heated mev-scale dark matter and the xenon1t electron recoil
  excess},}\ } (\bibinfo {year} {2020}),\ \Eprint
  {http://arxiv.org/abs/2006.12447} {arXiv:2006.12447 [hep-ph]} \BibitemShut
  {NoStop}%
\bibitem [{\citenamefont {Du}\ \emph {et~al.}(2020)\citenamefont {Du},
  \citenamefont {Liang}, \citenamefont {Liu}, \citenamefont {Tran},\ and\
  \citenamefont {Xue}}]{Du:2020ybt}%
  \BibitemOpen
  \bibfield  {author} {\bibinfo {author} {\bibfnamefont {M.}~\bibnamefont
  {Du}}, \bibinfo {author} {\bibfnamefont {J.}~\bibnamefont {Liang}}, \bibinfo
  {author} {\bibfnamefont {Z.}~\bibnamefont {Liu}}, \bibinfo {author}
  {\bibfnamefont {V.~Q.}\ \bibnamefont {Tran}}, \ and\ \bibinfo {author}
  {\bibfnamefont {Y.}~\bibnamefont {Xue}},\ }\href@noop {} {\  (\bibinfo {year}
  {2020})},\ \Eprint {http://arxiv.org/abs/2006.11949} {arXiv:2006.11949
  [hep-ph]} \BibitemShut {NoStop}%
\bibitem [{\citenamefont {Su}\ \emph {et~al.}(2020)\citenamefont {Su},
  \citenamefont {Wang}, \citenamefont {Wu}, \citenamefont {Yang},\ and\
  \citenamefont {Zhu}}]{Su:2020zny}%
  \BibitemOpen
  \bibfield  {author} {\bibinfo {author} {\bibfnamefont {L.}~\bibnamefont
  {Su}}, \bibinfo {author} {\bibfnamefont {W.}~\bibnamefont {Wang}}, \bibinfo
  {author} {\bibfnamefont {L.}~\bibnamefont {Wu}}, \bibinfo {author}
  {\bibfnamefont {J.~M.}\ \bibnamefont {Yang}}, \ and\ \bibinfo {author}
  {\bibfnamefont {B.}~\bibnamefont {Zhu}},\ }\href@noop {} {\  (\bibinfo {year}
  {2020})},\ \Eprint {http://arxiv.org/abs/2006.11837} {arXiv:2006.11837
  [hep-ph]} \BibitemShut {NoStop}%
\bibitem [{\citenamefont {Harigaya}\ \emph {et~al.}(2020)\citenamefont
  {Harigaya}, \citenamefont {Nakai},\ and\ \citenamefont
  {Suzuki}}]{Harigaya:2020ckz}%
  \BibitemOpen
  \bibfield  {author} {\bibinfo {author} {\bibfnamefont {K.}~\bibnamefont
  {Harigaya}}, \bibinfo {author} {\bibfnamefont {Y.}~\bibnamefont {Nakai}}, \
  and\ \bibinfo {author} {\bibfnamefont {M.}~\bibnamefont {Suzuki}},\
  }\href@noop {} {\  (\bibinfo {year} {2020})},\ \Eprint
  {http://arxiv.org/abs/2006.11938} {arXiv:2006.11938 [hep-ph]} \BibitemShut
  {NoStop}%
\bibitem [{\citenamefont {Paz}\ \emph {et~al.}(2020)\citenamefont {Paz},
  \citenamefont {Petrov}, \citenamefont {Tammaro},\ and\ \citenamefont
  {Zupan}}]{paz2020shining}%
  \BibitemOpen
  \bibfield  {author} {\bibinfo {author} {\bibfnamefont {G.}~\bibnamefont
  {Paz}}, \bibinfo {author} {\bibfnamefont {A.~A.}\ \bibnamefont {Petrov}},
  \bibinfo {author} {\bibfnamefont {M.}~\bibnamefont {Tammaro}}, \ and\
  \bibinfo {author} {\bibfnamefont {J.}~\bibnamefont {Zupan}},\ }\href@noop {}
  {\enquote {\bibinfo {title} {Shining dark matter in xenon1t},}\ } (\bibinfo
  {year} {2020}),\ \Eprint {http://arxiv.org/abs/2006.12462} {arXiv:2006.12462
  [hep-ph]} \BibitemShut {NoStop}%
\bibitem [{\citenamefont {Bell}\ \emph {et~al.}(2020)\citenamefont {Bell},
  \citenamefont {Dent}, \citenamefont {Dutta}, \citenamefont {Ghosh},
  \citenamefont {Kumar},\ and\ \citenamefont {Newstead}}]{bell2020explaining}%
  \BibitemOpen
  \bibfield  {author} {\bibinfo {author} {\bibfnamefont {N.~F.}\ \bibnamefont
  {Bell}}, \bibinfo {author} {\bibfnamefont {J.~B.}\ \bibnamefont {Dent}},
  \bibinfo {author} {\bibfnamefont {B.}~\bibnamefont {Dutta}}, \bibinfo
  {author} {\bibfnamefont {S.}~\bibnamefont {Ghosh}}, \bibinfo {author}
  {\bibfnamefont {J.}~\bibnamefont {Kumar}}, \ and\ \bibinfo {author}
  {\bibfnamefont {J.~L.}\ \bibnamefont {Newstead}},\ }\href@noop {} {\enquote
  {\bibinfo {title} {Explaining the xenon1t excess with luminous dark
  matter},}\ } (\bibinfo {year} {2020}),\ \Eprint
  {http://arxiv.org/abs/2006.12461} {arXiv:2006.12461 [hep-ph]} \BibitemShut
  {NoStop}%
\bibitem [{\citenamefont {Boehm}\ \emph {et~al.}(2020)\citenamefont {Boehm},
  \citenamefont {Cerdeno}, \citenamefont {Fairbairn}, \citenamefont {Machado},\
  and\ \citenamefont {Vincent}}]{Boehm:2020ltd}%
  \BibitemOpen
  \bibfield  {author} {\bibinfo {author} {\bibfnamefont {C.}~\bibnamefont
  {Boehm}}, \bibinfo {author} {\bibfnamefont {D.~G.}\ \bibnamefont {Cerdeno}},
  \bibinfo {author} {\bibfnamefont {M.}~\bibnamefont {Fairbairn}}, \bibinfo
  {author} {\bibfnamefont {P.~A.}\ \bibnamefont {Machado}}, \ and\ \bibinfo
  {author} {\bibfnamefont {A.~C.}\ \bibnamefont {Vincent}},\ }\href@noop {} {\
  (\bibinfo {year} {2020})},\ \Eprint {http://arxiv.org/abs/2006.11250}
  {arXiv:2006.11250 [hep-ph]} \BibitemShut {NoStop}%
\bibitem [{\citenamefont {Sierra}\ \emph {et~al.}(2020)\citenamefont {Sierra},
  \citenamefont {Romeri}, \citenamefont {Flores},\ and\ \citenamefont
  {Papoulias}}]{sierra2020light}%
  \BibitemOpen
  \bibfield  {author} {\bibinfo {author} {\bibfnamefont {D.~A.}\ \bibnamefont
  {Sierra}}, \bibinfo {author} {\bibfnamefont {V.~D.}\ \bibnamefont {Romeri}},
  \bibinfo {author} {\bibfnamefont {L.~J.}\ \bibnamefont {Flores}}, \ and\
  \bibinfo {author} {\bibfnamefont {D.~K.}\ \bibnamefont {Papoulias}},\
  }\href@noop {} {\enquote {\bibinfo {title} {Light vector mediators facing
  xenon1t data},}\ } (\bibinfo {year} {2020}),\ \Eprint
  {http://arxiv.org/abs/2006.12457} {arXiv:2006.12457 [hep-ph]} \BibitemShut
  {NoStop}%
\bibitem [{\citenamefont {D'Eramo}\ and\ \citenamefont
  {Thaler}(2010)}]{DEramo:2010keq}%
  \BibitemOpen
  \bibfield  {author} {\bibinfo {author} {\bibfnamefont {F.}~\bibnamefont
  {D'Eramo}}\ and\ \bibinfo {author} {\bibfnamefont {J.}~\bibnamefont
  {Thaler}},\ }\href {\doibase 10.1007/JHEP06(2010)109} {\bibfield  {journal}
  {\bibinfo  {journal} {JHEP}\ }\textbf {\bibinfo {volume} {06}},\ \bibinfo
  {pages} {109} (\bibinfo {year} {2010})},\ \Eprint
  {http://arxiv.org/abs/1003.5912} {arXiv:1003.5912 [hep-ph]} \BibitemShut
  {NoStop}%
\bibitem [{\citenamefont {Kamada}\ \emph {et~al.}(2018)\citenamefont {Kamada},
  \citenamefont {Kim}, \citenamefont {Kim},\ and\ \citenamefont
  {Sekiguchi}}]{Kamada:2017gfc}%
  \BibitemOpen
  \bibfield  {author} {\bibinfo {author} {\bibfnamefont {A.}~\bibnamefont
  {Kamada}}, \bibinfo {author} {\bibfnamefont {H.~J.}\ \bibnamefont {Kim}},
  \bibinfo {author} {\bibfnamefont {H.}~\bibnamefont {Kim}}, \ and\ \bibinfo
  {author} {\bibfnamefont {T.}~\bibnamefont {Sekiguchi}},\ }\href {\doibase
  10.1103/PhysRevLett.120.131802} {\bibfield  {journal} {\bibinfo  {journal}
  {Phys. Rev. Lett.}\ }\textbf {\bibinfo {volume} {120}},\ \bibinfo {pages}
  {131802} (\bibinfo {year} {2018})},\ \Eprint
  {http://arxiv.org/abs/1707.09238} {arXiv:1707.09238 [hep-ph]} \BibitemShut
  {NoStop}%
\bibitem [{\citenamefont {Smirnov}\ and\ \citenamefont
  {Beacom}(2020)}]{Smirnov:2020zwf}%
  \BibitemOpen
  \bibfield  {author} {\bibinfo {author} {\bibfnamefont {J.}~\bibnamefont
  {Smirnov}}\ and\ \bibinfo {author} {\bibfnamefont {J.~F.}\ \bibnamefont
  {Beacom}},\ }\href@noop {} {\  (\bibinfo {year} {2020})},\ \Eprint
  {http://arxiv.org/abs/2002.04038} {arXiv:2002.04038 [hep-ph]} \BibitemShut
  {NoStop}%
\bibitem [{\citenamefont {Hooper}(2019)}]{Hooper:2018kfv}%
  \BibitemOpen
  \bibfield  {author} {\bibinfo {author} {\bibfnamefont {D.}~\bibnamefont
  {Hooper}},\ }\href@noop {} {\bibfield  {journal} {\bibinfo  {journal} {PoS}\
  }\textbf {\bibinfo {volume} {TASI2018}},\ \bibinfo {pages} {010} (\bibinfo
  {year} {2019})},\ \Eprint {http://arxiv.org/abs/1812.02029} {arXiv:1812.02029
  [hep-ph]} \BibitemShut {NoStop}%
\bibitem [{\citenamefont {Roberts}\ and\ \citenamefont
  {Flambaum}(2019)}]{Roberts:2019chv}%
  \BibitemOpen
  \bibfield  {author} {\bibinfo {author} {\bibfnamefont {B.}~\bibnamefont
  {Roberts}}\ and\ \bibinfo {author} {\bibfnamefont {V.}~\bibnamefont
  {Flambaum}},\ }\href {\doibase 10.1103/PhysRevD.100.063017} {\bibfield
  {journal} {\bibinfo  {journal} {Phys. Rev. D}\ }\textbf {\bibinfo {volume}
  {100}},\ \bibinfo {pages} {063017} (\bibinfo {year} {2019})},\ \Eprint
  {http://arxiv.org/abs/1904.07127} {arXiv:1904.07127 [hep-ph]} \BibitemShut
  {NoStop}%
\bibitem [{\citenamefont {Riordan}\ \emph {et~al.}(1987)\citenamefont {Riordan}
  \emph {et~al.}}]{PhysRevLett.59.755}%
  \BibitemOpen
  \bibfield  {author} {\bibinfo {author} {\bibfnamefont {E.~M.}\ \bibnamefont
  {Riordan}} \emph {et~al.} (\bibinfo {collaboration} {E141}),\ }\href
  {\doibase 10.1103/PhysRevLett.59.755} {\bibfield  {journal} {\bibinfo
  {journal} {Phys. Rev. Lett.}\ }\textbf {\bibinfo {volume} {59}},\ \bibinfo
  {pages} {755} (\bibinfo {year} {1987})}\BibitemShut {NoStop}%
\bibitem [{\citenamefont {Banerjee}\ \emph {et~al.}(2020)\citenamefont
  {Banerjee} \emph {et~al.}}]{Banerjee:2019hmi}%
  \BibitemOpen
  \bibfield  {author} {\bibinfo {author} {\bibfnamefont {D.}~\bibnamefont
  {Banerjee}} \emph {et~al.} (\bibinfo {collaboration} {NA64}),\ }\href
  {\doibase 10.1103/PhysRevD.101.071101} {\bibfield  {journal} {\bibinfo
  {journal} {Phys. Rev. D}\ }\textbf {\bibinfo {volume} {101}},\ \bibinfo
  {pages} {071101} (\bibinfo {year} {2020})},\ \Eprint
  {http://arxiv.org/abs/1912.11389} {arXiv:1912.11389 [hep-ex]} \BibitemShut
  {NoStop}%
\bibitem [{\citenamefont {Banerjee}\ \emph {et~al.}(2019)\citenamefont
  {Banerjee} \emph {et~al.}}]{NA64:2019imj}%
  \BibitemOpen
  \bibfield  {author} {\bibinfo {author} {\bibfnamefont {D.}~\bibnamefont
  {Banerjee}} \emph {et~al.},\ }\href {\doibase 10.1103/PhysRevLett.123.121801}
  {\bibfield  {journal} {\bibinfo  {journal} {Phys. Rev. Lett.}\ }\textbf
  {\bibinfo {volume} {123}},\ \bibinfo {pages} {121801} (\bibinfo {year}
  {2019})},\ \Eprint {http://arxiv.org/abs/1906.00176} {arXiv:1906.00176
  [hep-ex]} \BibitemShut {NoStop}%
\bibitem [{\citenamefont {Lees}\ \emph {et~al.}(2014)\citenamefont {Lees} \emph
  {et~al.}}]{Lees:2014xha}%
  \BibitemOpen
  \bibfield  {author} {\bibinfo {author} {\bibfnamefont {J.}~\bibnamefont
  {Lees}} \emph {et~al.} (\bibinfo {collaboration} {BaBar}),\ }\href {\doibase
  10.1103/PhysRevLett.113.201801} {\bibfield  {journal} {\bibinfo  {journal}
  {Phys. Rev. Lett.}\ }\textbf {\bibinfo {volume} {113}},\ \bibinfo {pages}
  {201801} (\bibinfo {year} {2014})},\ \Eprint {http://arxiv.org/abs/1406.2980}
  {arXiv:1406.2980 [hep-ex]} \BibitemShut {NoStop}%
\bibitem [{\citenamefont {Lees}\ \emph {et~al.}(2017)\citenamefont {Lees} \emph
  {et~al.}}]{Lees:2017lec}%
  \BibitemOpen
  \bibfield  {author} {\bibinfo {author} {\bibfnamefont {J.}~\bibnamefont
  {Lees}} \emph {et~al.} (\bibinfo {collaboration} {BaBar}),\ }\href {\doibase
  10.1103/PhysRevLett.119.131804} {\bibfield  {journal} {\bibinfo  {journal}
  {Phys. Rev. Lett.}\ }\textbf {\bibinfo {volume} {119}},\ \bibinfo {pages}
  {131804} (\bibinfo {year} {2017})},\ \Eprint
  {http://arxiv.org/abs/1702.03327} {arXiv:1702.03327 [hep-ex]} \BibitemShut
  {NoStop}%
\bibitem [{\citenamefont {Tanabashi}\ \emph {et~al.}(2018)\citenamefont
  {Tanabashi} \emph {et~al.}}]{PhysRevD.98.030001}%
  \BibitemOpen
  \bibfield  {author} {\bibinfo {author} {\bibfnamefont {M.}~\bibnamefont
  {Tanabashi}} \emph {et~al.} (\bibinfo {collaboration} {Particle Data
  Group}),\ }\href {\doibase 10.1103/PhysRevD.98.030001} {\bibfield  {journal}
  {\bibinfo  {journal} {Phys. Rev. D}\ }\textbf {\bibinfo {volume} {98}},\
  \bibinfo {pages} {030001} (\bibinfo {year} {2018})}\BibitemShut {NoStop}%
\bibitem [{\citenamefont {Barate}\ \emph {et~al.}(2000)\citenamefont {Barate}
  \emph {et~al.}}]{Barate:1999qx}%
  \BibitemOpen
  \bibfield  {author} {\bibinfo {author} {\bibfnamefont {R.}~\bibnamefont
  {Barate}} \emph {et~al.} (\bibinfo {collaboration} {ALEPH}),\ }\href
  {\doibase 10.1007/s100529900223} {\bibfield  {journal} {\bibinfo  {journal}
  {Eur. Phys. J. C}\ }\textbf {\bibinfo {volume} {12}},\ \bibinfo {pages} {183}
  (\bibinfo {year} {2000})},\ \Eprint {http://arxiv.org/abs/hep-ex/9904011}
  {arXiv:hep-ex/9904011} \BibitemShut {NoStop}%
\bibitem [{\citenamefont {Alcaraz}\ \emph {et~al.}(2006)\citenamefont {Alcaraz}
  \emph {et~al.}}]{Alcaraz:2006mx}%
  \BibitemOpen
  \bibfield  {author} {\bibinfo {author} {\bibfnamefont {J.}~\bibnamefont
  {Alcaraz}} \emph {et~al.} (\bibinfo {collaboration} {ALEPH, DELPHI, L3, OPAL,
  LEP Electroweak Working Group}),\ }\href@noop {} {\  (\bibinfo {year}
  {2006})},\ \Eprint {http://arxiv.org/abs/hep-ex/0612034}
  {arXiv:hep-ex/0612034} \BibitemShut {NoStop}%
\bibitem [{\citenamefont {Freitas}\ \emph {et~al.}(2014)\citenamefont
  {Freitas}, \citenamefont {Lykken}, \citenamefont {Kell},\ and\ \citenamefont
  {Westhoff}}]{Freitas:2014pua}%
  \BibitemOpen
  \bibfield  {author} {\bibinfo {author} {\bibfnamefont {A.}~\bibnamefont
  {Freitas}}, \bibinfo {author} {\bibfnamefont {J.}~\bibnamefont {Lykken}},
  \bibinfo {author} {\bibfnamefont {S.}~\bibnamefont {Kell}}, \ and\ \bibinfo
  {author} {\bibfnamefont {S.}~\bibnamefont {Westhoff}},\ }\href {\doibase
  10.1007/JHEP09(2014)155} {\bibfield  {journal} {\bibinfo  {journal} {JHEP}\
  }\textbf {\bibinfo {volume} {05}},\ \bibinfo {pages} {145} (\bibinfo {year}
  {2014})},\ \bibinfo {note} {[Erratum: JHEP 09, 155 (2014)]},\ \Eprint
  {http://arxiv.org/abs/1402.7065} {arXiv:1402.7065 [hep-ph]} \BibitemShut
  {NoStop}%
\bibitem [{\citenamefont {Abdallah}\ \emph {et~al.}(2009)\citenamefont
  {Abdallah} \emph {et~al.}}]{Abdallah:2008aa}%
  \BibitemOpen
  \bibfield  {author} {\bibinfo {author} {\bibfnamefont {J.}~\bibnamefont
  {Abdallah}} \emph {et~al.} (\bibinfo {collaboration} {DELPHI}),\ }\href
  {\doibase 10.1140/epjc/s10052-009-0874-9} {\bibfield  {journal} {\bibinfo
  {journal} {Eur. Phys. J. C}\ }\textbf {\bibinfo {volume} {60}},\ \bibinfo
  {pages} {17} (\bibinfo {year} {2009})},\ \Eprint
  {http://arxiv.org/abs/0901.4486} {arXiv:0901.4486 [hep-ex]} \BibitemShut
  {NoStop}%
\bibitem [{\citenamefont {Fox}\ \emph {et~al.}(2011)\citenamefont {Fox},
  \citenamefont {Harnik}, \citenamefont {Kopp},\ and\ \citenamefont
  {Tsai}}]{Fox:2011fx}%
  \BibitemOpen
  \bibfield  {author} {\bibinfo {author} {\bibfnamefont {P.~J.}\ \bibnamefont
  {Fox}}, \bibinfo {author} {\bibfnamefont {R.}~\bibnamefont {Harnik}},
  \bibinfo {author} {\bibfnamefont {J.}~\bibnamefont {Kopp}}, \ and\ \bibinfo
  {author} {\bibfnamefont {Y.}~\bibnamefont {Tsai}},\ }\href {\doibase
  10.1103/PhysRevD.84.014028} {\bibfield  {journal} {\bibinfo  {journal} {Phys.
  Rev. D}\ }\textbf {\bibinfo {volume} {84}},\ \bibinfo {pages} {014028}
  (\bibinfo {year} {2011})},\ \Eprint {http://arxiv.org/abs/1103.0240}
  {arXiv:1103.0240 [hep-ph]} \BibitemShut {NoStop}%
\bibitem [{\citenamefont {Kaplinghat}\ \emph {et~al.}(2000)\citenamefont
  {Kaplinghat}, \citenamefont {Knox},\ and\ \citenamefont
  {Turner}}]{Kaplinghat:2000vt}%
  \BibitemOpen
  \bibfield  {author} {\bibinfo {author} {\bibfnamefont {M.}~\bibnamefont
  {Kaplinghat}}, \bibinfo {author} {\bibfnamefont {L.}~\bibnamefont {Knox}}, \
  and\ \bibinfo {author} {\bibfnamefont {M.~S.}\ \bibnamefont {Turner}},\
  }\href {\doibase 10.1103/PhysRevLett.85.3335} {\bibfield  {journal} {\bibinfo
   {journal} {Phys. Rev. Lett.}\ }\textbf {\bibinfo {volume} {85}},\ \bibinfo
  {pages} {3335} (\bibinfo {year} {2000})},\ \Eprint
  {http://arxiv.org/abs/astro-ph/0005210} {arXiv:astro-ph/0005210} \BibitemShut
  {NoStop}%
\bibitem [{\citenamefont {Beacom}\ \emph {et~al.}(2007)\citenamefont {Beacom},
  \citenamefont {Bell},\ and\ \citenamefont {Mack}}]{Beacom:2006tt}%
  \BibitemOpen
  \bibfield  {author} {\bibinfo {author} {\bibfnamefont {J.~F.}\ \bibnamefont
  {Beacom}}, \bibinfo {author} {\bibfnamefont {N.~F.}\ \bibnamefont {Bell}}, \
  and\ \bibinfo {author} {\bibfnamefont {G.~D.}\ \bibnamefont {Mack}},\ }\href
  {\doibase 10.1103/PhysRevLett.99.231301} {\bibfield  {journal} {\bibinfo
  {journal} {Phys. Rev. Lett.}\ }\textbf {\bibinfo {volume} {99}},\ \bibinfo
  {pages} {231301} (\bibinfo {year} {2007})},\ \Eprint
  {http://arxiv.org/abs/astro-ph/0608090} {arXiv:astro-ph/0608090} \BibitemShut
  {NoStop}%
\bibitem [{\citenamefont {Aprile}\ \emph {et~al.}(2019)\citenamefont {Aprile}
  \emph {et~al.}}]{Aprile:2019jmx}%
  \BibitemOpen
  \bibfield  {author} {\bibinfo {author} {\bibfnamefont {E.}~\bibnamefont
  {Aprile}} \emph {et~al.} (\bibinfo {collaboration} {XENON}),\ }\href
  {\doibase 10.1103/PhysRevLett.123.241803} {\bibfield  {journal} {\bibinfo
  {journal} {Phys. Rev. Lett.}\ }\textbf {\bibinfo {volume} {123}},\ \bibinfo
  {pages} {241803} (\bibinfo {year} {2019})},\ \Eprint
  {http://arxiv.org/abs/1907.12771} {arXiv:1907.12771 [hep-ex]} \BibitemShut
  {NoStop}%
\bibitem [{\citenamefont {Amole}\ \emph {et~al.}(2019)\citenamefont {Amole}
  \emph {et~al.}}]{Amole:2019fdf}%
  \BibitemOpen
  \bibfield  {author} {\bibinfo {author} {\bibfnamefont {C.}~\bibnamefont
  {Amole}} \emph {et~al.} (\bibinfo {collaboration} {PICO}),\ }\href {\doibase
  10.1103/PhysRevD.100.022001} {\bibfield  {journal} {\bibinfo  {journal}
  {Phys. Rev. D}\ }\textbf {\bibinfo {volume} {100}},\ \bibinfo {pages}
  {022001} (\bibinfo {year} {2019})},\ \Eprint
  {http://arxiv.org/abs/1902.04031} {arXiv:1902.04031 [astro-ph.CO]}
  \BibitemShut {NoStop}%
\bibitem [{\citenamefont {Gluscevic}\ and\ \citenamefont
  {Boddy}(2018)}]{Gluscevic:2017ywp}%
  \BibitemOpen
  \bibfield  {author} {\bibinfo {author} {\bibfnamefont {V.}~\bibnamefont
  {Gluscevic}}\ and\ \bibinfo {author} {\bibfnamefont {K.~K.}\ \bibnamefont
  {Boddy}},\ }\href {\doibase 10.1103/PhysRevLett.121.081301} {\bibfield
  {journal} {\bibinfo  {journal} {Phys. Rev. Lett.}\ }\textbf {\bibinfo
  {volume} {121}},\ \bibinfo {pages} {081301} (\bibinfo {year} {2018})},\
  \Eprint {http://arxiv.org/abs/1712.07133} {arXiv:1712.07133 [astro-ph.CO]}
  \BibitemShut {NoStop}%
\bibitem [{\citenamefont {Xu}\ \emph {et~al.}(2018)\citenamefont {Xu},
  \citenamefont {Dvorkin},\ and\ \citenamefont {Chael}}]{Xu:2018efh}%
  \BibitemOpen
  \bibfield  {author} {\bibinfo {author} {\bibfnamefont {W.~L.}\ \bibnamefont
  {Xu}}, \bibinfo {author} {\bibfnamefont {C.}~\bibnamefont {Dvorkin}}, \ and\
  \bibinfo {author} {\bibfnamefont {A.}~\bibnamefont {Chael}},\ }\href
  {\doibase 10.1103/PhysRevD.97.103530} {\bibfield  {journal} {\bibinfo
  {journal} {Phys. Rev. D}\ }\textbf {\bibinfo {volume} {97}},\ \bibinfo
  {pages} {103530} (\bibinfo {year} {2018})},\ \Eprint
  {http://arxiv.org/abs/1802.06788} {arXiv:1802.06788 [astro-ph.CO]}
  \BibitemShut {NoStop}%
\bibitem [{\citenamefont {Slatyer}\ and\ \citenamefont
  {Wu}(2018)}]{Slatyer:2018aqg}%
  \BibitemOpen
  \bibfield  {author} {\bibinfo {author} {\bibfnamefont {T.~R.}\ \bibnamefont
  {Slatyer}}\ and\ \bibinfo {author} {\bibfnamefont {C.-L.}\ \bibnamefont
  {Wu}},\ }\href {\doibase 10.1103/PhysRevD.98.023013} {\bibfield  {journal}
  {\bibinfo  {journal} {Phys. Rev. D}\ }\textbf {\bibinfo {volume} {98}},\
  \bibinfo {pages} {023013} (\bibinfo {year} {2018})},\ \Eprint
  {http://arxiv.org/abs/1803.09734} {arXiv:1803.09734 [astro-ph.CO]}
  \BibitemShut {NoStop}%
\bibitem [{\citenamefont {Essig}\ \emph {et~al.}(2013)\citenamefont {Essig}
  \emph {et~al.}}]{Essig:2013lka}%
  \BibitemOpen
  \bibfield  {author} {\bibinfo {author} {\bibfnamefont {R.}~\bibnamefont
  {Essig}} \emph {et~al.},\ }in\ \href@noop {} {\emph {\bibinfo {booktitle}
  {{Community Summer Study 2013}: {Snowmass on the Mississippi}}}}\ (\bibinfo
  {year} {2013})\ \Eprint {http://arxiv.org/abs/1311.0029} {arXiv:1311.0029
  [hep-ph]} \BibitemShut {NoStop}%
\end{thebibliography}%

\end{document}